\documentclass{elsarticle}                 
\pdfoutput=1
\usepackage{epsfig}
\usepackage{graphicx}
\usepackage{epstopdf}
\usepackage{amssymb}
\usepackage{wasysym}
\usepackage{color}
\usepackage{lineno}
\RequirePackage{xspace}

          \usepackage{graphicx}%

\begin{document}

\begin{frontmatter}

  \title{The preamplifier-shaper for the ALICE TPC-Detector }

  \author[1]{H. K. Soltveit\footnote{Corresponding author, e-mail:soltveit@physi.uni-heidelberg.de}}
  \author[1]{J. Stachel}
  \author[2]{P. Braun-Munzinger}
  \author[3]{L. Musa}
  \author[4]{H, A. Gustafsson$^\dagger$}
  \author[5]{U. Bonnes}
  \author[5]{H. Oeschler}
  \author[4]{L. Osterman}
  \author[5]{S. Lang}
  \author{for the ALICE TPC Collaboration}

  \address[1]{Physikalisches Institut, Ruprecht-Karls-Universit\"{a}t Heidelberg, Heidelberg, Germany}
  \address[2]{GSI Helmholtzzentrum f\"{u}r Schwerionenforschung GmbH, Darmstadt, Germany}
  \address[3]{European Organization for Nuclear Research (CERN), Geneva}
   \address[4]{Division of Experimental High Energy Physics, University of Lund, Lund, Sweden}
   \address[5]{Institut f\"{u}r Kernphysik, Technische Universit\"{a}t Darmstadt, Darmstadt, Germany}
  
  \date{today}

\begin{abstract}
In this paper the PreAmplifier ShAper (PASA) for the Time Projection
Chamber (TPC) of the ALICE experiment at LHC is presented. The ALICE TPC PASA is an ASIC that integrates 
16 identical channels, each consisting of Charge Sensitive Amplifiers (CSA) followed by a Pole-Zero network,
self-adaptive bias network, two second-order bridged-T filters, two
non-inverting level shifters and a start-up circuit. The circuit is
optimized for a detector capacitance of 18-25 pF. For an input capacitance of 25 pF, the PASA features a conversion gain of  12.74~mV/fC, a peaking time of 160 ns, a FWHM  of 190 ns, a power consumption of 11.65 mW/ch and an equivalent noise charge of 244e + 17e/pF. The circuit recovers smoothly to the baseline
in about 600 ns.  An integral
non-linearity of 0.19\%  with an output swing of about 2.1 V is also achieved. 
The total area of the chip is 18 mm$^2$ and is implemented in AMS's C35B3C1 0.35 micron
CMOS technology.  Detailed characterization test were performed on about 48000 PASA 
circuits before mounting them on the ALICE TPC front-end cards. After more than two years of operation of the ALICE TPC with p-p and Pb-Pb collisions, the PASA has demonstrated to fulfill all requirements.

\end{abstract}

\end{frontmatter}

\section{Introduction}

The Time Projection Chamber (TPC) is the main tracking detector in the
central barrel of the ALICE experiment \cite{alice} at the
CERN LHC. It is a 90 m$^3$ cylinder filled with gas and divided in two
drift regions by the central electrode located at its axial center. A
field cage creates a uniform electric field along each half of the
chamber.

Charged particles traversing the TPC volume ionize the gas
along their paths liberating electrons that drift toward the end plates
of the chamber. The necessary signal amplification is provided through
avalanche effects in the vicinity of the anode wires. Moving from the
anode wire toward the surrounding electrodes, the positive ions
created in the avalanche induce a positive current signal which
characterized by a fast rise time (less than 1 ns) and a long tail
with a rather complex shape. It carries a charge that for the minimum
ionizing particle that is in the order of 4.8~fC. The readout of the signal is done by
the 557 568 pads that form the cathode plane of the conventional wire
chambers located at the TPC end caps.
The amplitude, which is different for the different pad sizes, has a 
typical value of 7~$\mu$A. The signal is delivered on the detector
impedance which, to a very good approximation, is a pure 
capacitance of the order of 12~pF.

The signals from the pads are passed to 4356 front-end cards, located about 10 cm away from the pad plane, via flexible kapton cables.
In the front-end card the charge-sensitive shaping amplifier
of the PASA circuit transforms the charge induced in the pads into a differential
semi-Gaussian signal that is fed to the input of the ALTRO chip \cite{ALTRO}. 

This paper addresses the design of the PASA that is fabricated in a 0.35 micron CMOS
technology. In section \ref{sec:over}, the used PASA architecture is presented. This
architecture consists of several blocks that are described in
sections \ref{sec:coramp}, \ref{sec:shaper}, \ref{sec:nonin}, and \ref{sec:bias}. 
In section~\ref{sec:layout} the layout techniques are explained.
In section~\ref{sec:simu}, the performance of the chip at the
simulation level is shown. The standalone test results of all produced
chips and their performance in the final system are described in
section~\ref{sec:meassys}. The conclusions are presented in 
section~\ref{sec:conclu}.

\section{ALICE TPC PASA overview}
\label{sec:over}

\begin{figure}[hb]
\begin{center}
  \includegraphics[width=0.9\textwidth]{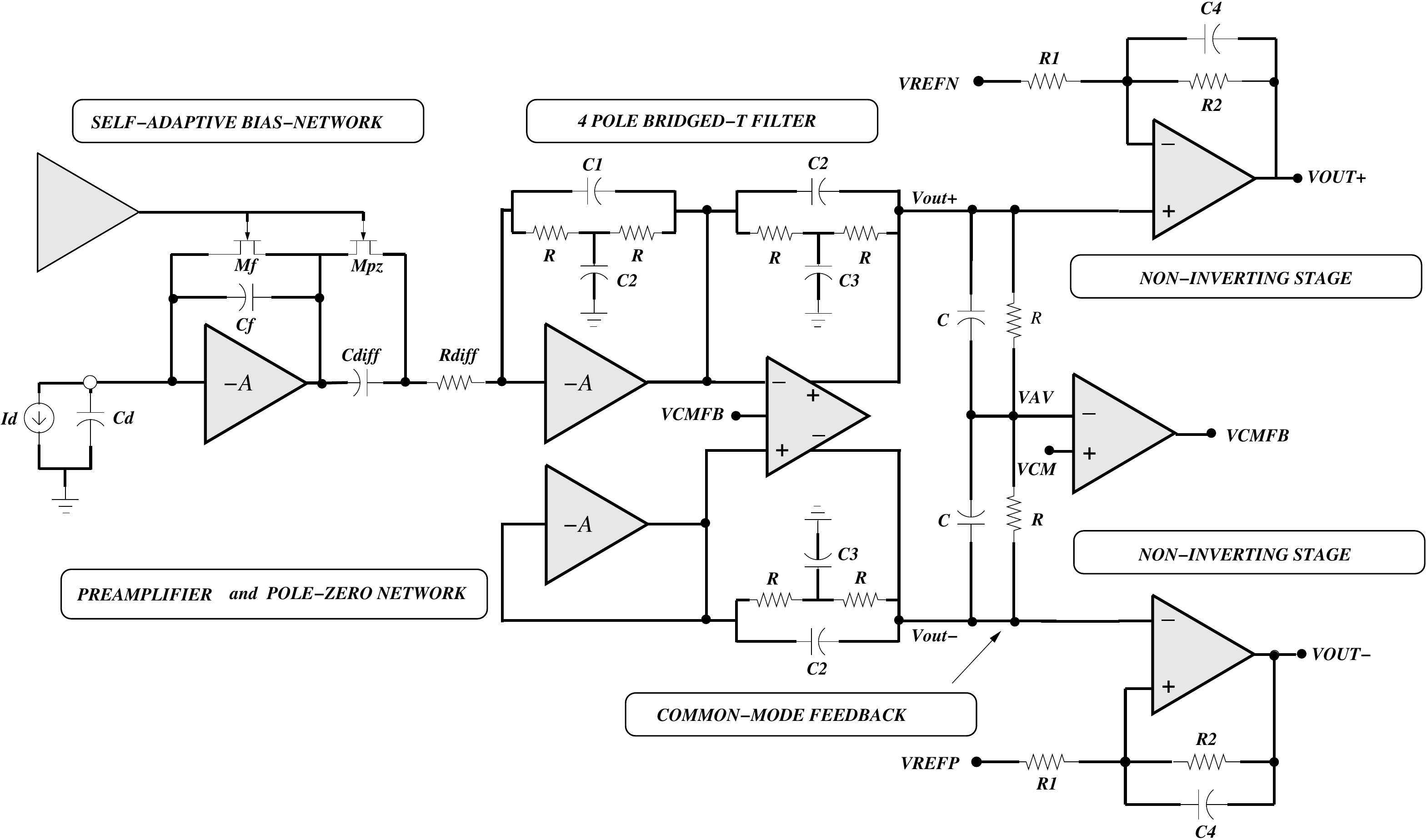}
\caption{A simplified block diagram of the PreAmplifier-ShAper (PASA)
  signal processing chain.}
\label{fig:blockpasa}
\end{center}
\end{figure}

The list of requirements given as guideline prior to the design of the PASA
is presented in Table~\ref{tab:req}. A simplified block diagram of the PASA signal processing chain
developed in order to fulfill them is shown in Fig.~\ref{fig:blockpasa}. 
\begin{table}[htb]
\begin{center}
\begin{tabular}{|c||c|}\hline
    Parameter              & Specification    \\ \hline \hline
    Noise                  &$<$ 1000e     \\ \hline
    Shaping time (ns)      & 190          \\ \hline
    Non-linearity          &$<$ 1\%       \\ \hline
    Crosstalk              &$<$ 0.3\%     \\ \hline
    Total max capacitance (pF) & 25        \\ \hline
    Conv. gain (mV/fC)     & 12           \\ \hline
    Power con.(mW/ch)      & $<$ 20       \\ \hline
\end{tabular}
\end{center}
\caption{List of requirements given for the design of the ALICE TPC PASA.}
\label{tab:req}
\end{table}
The ASIC has 16 equal channels, consisting of a ESD network made of six diodes and a 
50~$\Omega$ resistor, a positive polarity Charge Sensitive Amplifier (CSA) with a capacitive
feedback Cf and a resistive feedback Rf(Mf) connected in parallel, a Pole-Zero
Cancellation (PZC) network, a self-adaptive bias network, a CR filter and two (RC)$^2$ bridged-T
filters, a Common-Mode Feed-Back network (CMFB) and two quasi-differential gain-2 
amplifiers. Not shown in the block diagram is the internal bias-network.

Since the charge Qin delivered by the TPC detector readout chambers is very small (typically 7$\mu$A),  and short (1ns), it is  unsuitable for immediate
signal processing. Therefore, the input signal is first integrated and amplified by the
CSA producing at its output a voltage signal Vout, whose amplitude is
proportional to the total charge Qin and characterized by a long decay
time constant, which is defined by the feedback network parameters  ($\tau$ =Rf(Mf)*Cf= 4.2 $\mu$s) of
the CSA:
\begin{equation}
Vout =\frac{Qin}{Cf} \cdot\exp{(-t/\tau)}
\label{eq:qin}
\end{equation}
As seen in Fig.~\ref{fig:blockpasa} and mathematically represented in
Eq.~\ref{eq:qin} a NMOS transistor Rf(Mf) operated in the
subthreshold region is connected in parallel to the feedback
capacitor Cf. The purpose of this transistor is to avoid saturation of
the CSA, by continuously discharging the feedback capacitance Cf. 
As later explained in section 3, this transistor will contribute to the parallel noise at the CSA
input. Therefore, in order to limit its noise contribution, it was
chosen to be a high value (10M$\Omega$) as a compromise between noise and count rate. 
The relatively big resistance value chosen here, was implemented as an
active feedback transistor instead of as a passive resistor, that otherwise would
have increased the area, parasitic capacitances, speed and noise.
 
To operate a transistor with such a high resistance makes its drain-to-source resistance Rds
sensitive to process, temperature and bias conditions. To neutralize this sensitivity the self-adaptive
bias-network developed by \cite{biasnet} was adopted. This scheme has the advantage that it 
keeps the difference Vg(Mf)-Vin(M1) constant, making Rf(Mf) independent of the effects mentioned above.

Still, the relatively long discharge time constant of the CSA makes it vulnerable
to pile-up. The low frequency part of the pulse is then removed by the
C$_{diff}$R$_{diff}$ -filter stage. 
Due to the exponential decay of the CSA feedback network in
combination with the differentiator network, an undershoot is created
at the shaper output with the same time constant as the CSA of $\tau$ = Rf(Mf)*Cf. 
This undershoot was removed by creating a PZC circuit by
adding a transistor Rpz (Mpz) in parallel to the capacitor Cdiff in
the differentiator stage. This creates a Zero in the transfer function that  cancels the low frequency pole introduced by
the CSA feedback network. In order to ensure that the Zero introduced
by the network Mpz-C$_{diff}$ adapts dynamically and accurately cancels the
pole associated with the CSA feedback network Cf-Mf(Rf), the gate 
voltage of the transistor Mpz(Rpz) is controlled by the same
self-adaptive bias-network as the Mf(Rf) transistor.

To improve the linearity of the charge to voltage gain, a voltage divider
was inserted between the output of the preamplifier and the source of Mf \cite{2}. 
This reduces the variation for Rf(Mf) during the operation. 
Moreover, t$_{decay}$ = k*Rf(Mf)*Cf is increased by the factor k = (R1+R2)/R2, resulting in a very precise return to 
the baseline. Owing to the method described above, the maximum undershoot created by the mismatch of the components due to the variation of the CMOS process parameters, never exceed the 0.1\% of the
circuit dynamic range.

The output signal of the CSA and the PZC network is then amplified and shaped by two 2$^{nd}$
order bridged-T filters (see section~\ref{sec:shaper}) to optimize the
signal-to-noise ratio (SNR) and to limit the signal bandwidth. 

The shaped signal is then fed to the last stage consisting of two
non-inverting stages (see section~\ref{sec:nonin}), built around a Miller operational amplifier,
each providing a fixed gain of 2.

To adapt to the ALTRO input dynamic range, the two output DC levels
VOUT+ and VOUT- are defined by three externally given references, 
VREFN (0.56~V), VREFP (1.056~V) and VCM (1.056~V).

\section{The core amplifier} 
\label{sec:coramp}

\begin{figure}[hbt]

\begin{center}
\includegraphics[width=0.75\textwidth]{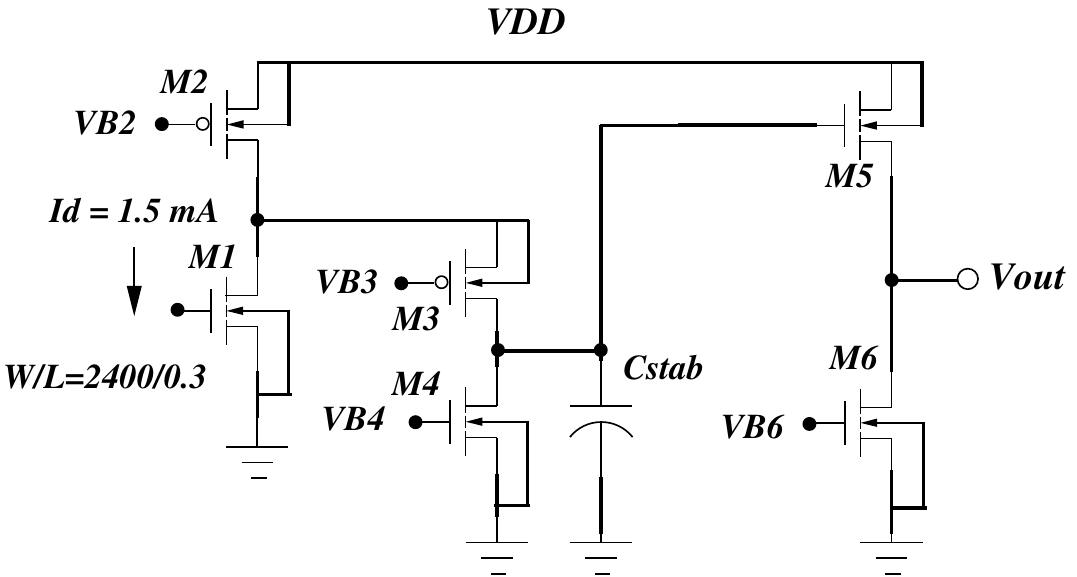} 

\caption{A schematic diagram of the Charge Sensitive Amplifier (CSA). }
\label{fig:CSA}
\end{center}
\end{figure}

The core amplifier topology and the input transistor (PMOS
versus NMOS) are key choices in the design of a low-power low-noise charge sensitive amplifier, since they typically represent the largest contribution to the total power dissipation and the overall system noise. The chosen topology
(Fig.~\ref{fig:CSA}) is based on a single-ended folded cascode amplifier
followed by a source follower. Owing to the short shaping time (190 ns), the white noise is the dominating source. Therefore, an NMOS transistor was the natural choice, since it gives a higher 
transconductance $gm$ and therefore better noise than a PMOS 
transistor under equal conditions.
  
The CSA has been optimized for a detector capacitance of 25 pF. 
The chosen aspect ratio (W/L)  for the NMOS input
transistor M1 of 2400/0.3 were based on the theoretical calculations
described in subsection \ref{subsec:theo}. 

\begin{figure}[hbt]
\begin{center}

\includegraphics[width=0.9\textwidth]{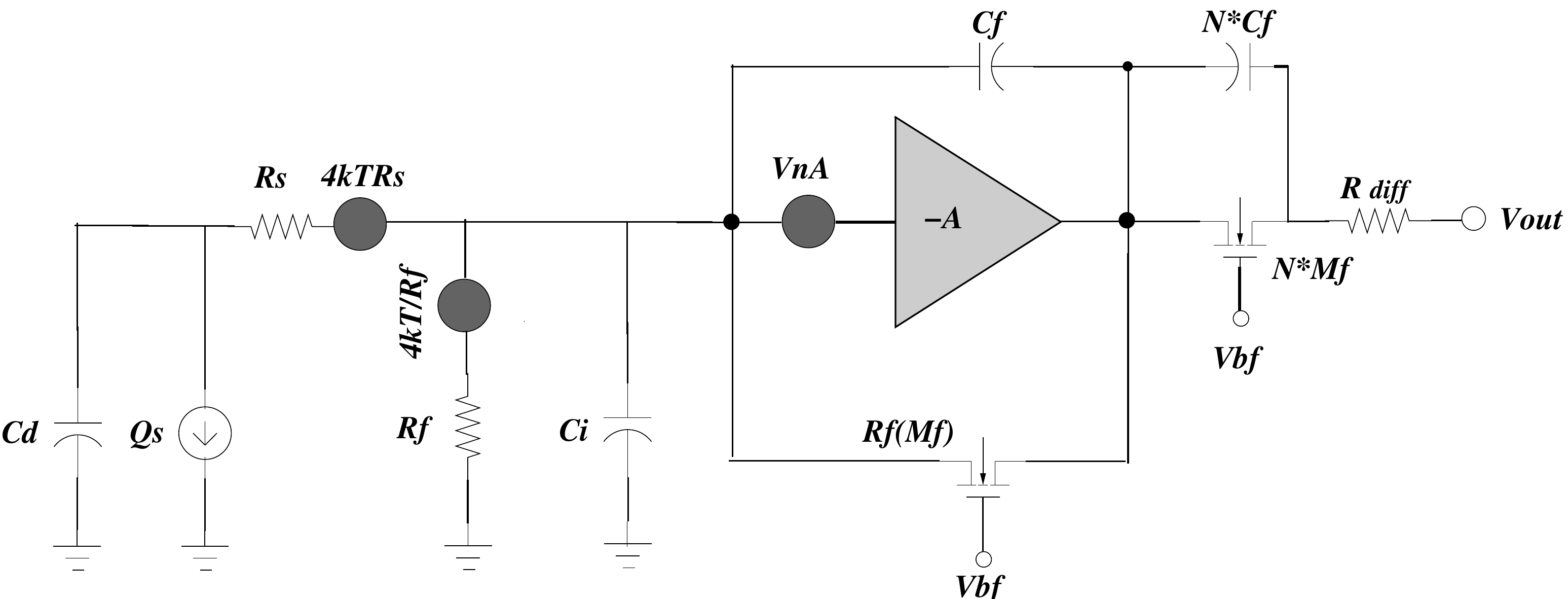}

\caption{Block schematic of the CSA including pole-zero network and dynamic capacitance Ci.}
\label{fig:CSA2}
\end{center}
\end{figure}

To avoid cross-talk between adjacent channels, which can deteriorate the
energy and position resolution,  the input impedance (Z$_{in}$) is
made as small as possible. In this case it is calculated by using:
\begin{equation}
Z_{in} = \frac{C_{out}}{{gm}\cdot C_{f}}
\end{equation}
where the feedback capacitance $C_f$ is 0.42 pF, the
capacitance $C_{out}$ of the dominant pole is 1.4~pF and the 
input of the amplifier has a $gm$ of 33 mS.
This gives an effective input impedance of about 100~$\Omega$.

The open-loop gain of this stage is 80 dB and the power consumption is about 6~mW.

\subsection{Noise optimization of the PASA}
\label{subsec:theo}

In order to fulfill the requirement of a signal-to-noise ratio of 30:1 for the signal of 4.8fC produced by a MIP, the PASA has to feature a maximum noise of 1000 e.
Noise here is referred to as equivalent noise Charge (ENC) which can
directly be compared to the number of
electrons created in the detector at the input of the CSA.

The noise model circuit for the PASA is shown in Fig.~\ref{fig:CSA2}, 
where the three most important noise sources are represented as red dots.
These are the series thermal noise ($4kTRs$) from the ESD protection circuit
(70~$\Omega$), the parasitic bulk resistance noise, the gate
resistance thermal noise, the thermal channel noise ($VnA$) from the
input transistor M1, the flicker noise from the input
transistor M1 and the parallel thermal noise contribution ($4kT/Rf$) from the feedback
transistor R$_{f}$(M$_{f}$).  

From the given specifications four main parameters are fixed: 
the shaping time $\tau$, the total detector capacitance C$_{T}$, 
the noise at 25~pF, the filter order (n=4), and the power consumption. In
addition, a series resistance of 70~$\Omega$ in the ESD network was chosen.

Therefore, to find the optimum ENC$_{T}$, the minimum gate length of
transistor M1 available from the technology was chosen to achieve a high $gm$, a low thermal
noise, a low power and a high speed.
Given the parameters mentioned above, the only free parameter left is the width W of the input transistor M1. 

Since the PASA performs a CR-RC$^4$ semi-Gaussian shaping, the relative
ENC contributions from the four dominating and uncorrelated noise sources can then be
expressed as~\cite{sansenchang}:
 
\begin{equation}
ENC^{2}_{R_{S}} = {4kT}{R_{S}}\frac{C^{2}_{T}B(\frac{3}{2},n-\frac{1}{2})n}{4\pi \tau_{s}}(\frac{n!e^{2n}}{n^{2n}})
\label{eq:rs}
\end{equation}

\begin{equation}
ENC^{2}_{R_{fb}} = \frac{4kT}{R_{fb}}\tau_{s}\frac{B(\frac{3}{2},n-\frac{1}{2})n}{4\pi}(\frac{n!e^{2n}}{n^{2n}})
\label{eq:rfb}
\end{equation}

\begin{equation}
ENC^{2}_{th} = \frac{8}{3}kT\frac{1}{gm}\frac{C^{2}_{T}B(\frac{3}{2},n-\frac{1}{2})n}{q^{2}4\pi\tau_{s}}(\frac{n!^{2}e^{2n}}{n^{2n}})
\label{eq:th}
\end{equation}

\begin{equation}
ENC^{2}_{\frac{1}{f}} = \frac{K_{f}}{C^{2}_{ox}WL}\frac{C^{2}_{T}}{q^{2}2n}(\frac{n!^{2}e^{2n}}{n^{2n}})
\label{eq:overf}
\end{equation}

where $k$ is Boltzmann constant, $T$ is the temperature (Kelvin), $gm$ is the 
transconductance for the input transistor (mS), $K_f$ is (Flicker noise
coefficient),
$B(X,Y)$ is the  Beta function~\cite{sansenchang}, and $C_{T}$ is the total
capacitance at the input of M1. $ C_{T}=C_{p}+C_{f}+C_{gs}+C_{gd}+C_{FE-board}$, where
$C_{p}$ is the pad capacitance, $C_{f}$ is the feedback capacitance, $C_{gs}$ and $C_{gd}$ are the
gate-source and gate-drain capacitance of M1, respectively, and
$C_{FE-board}$ is the capacitance of the traces on the board.

Eq.~(\ref{eq:rs}) represents the noise contribution from the series resistance of the ESD
network and its parasitic resistances. The parasitic resistances
include the resistive poly-gate and the
distributed substrate resistance. The main contributor here is the
current limited resistor R$_{s}$ of 70~$\Omega$ implemented in the
ESD network. Its size was chosen as a compromise between noise and protection 
capability. This resistor produces a thermal noise voltage that is converted to a
``noise charge'' which at the output of the CSA gives a signal
proportional to the charge flowing into the CSA. 
As mentioned above, the parasitic resistances related to the resistive poly-gate and the distributed
substrate resistance also have an impact on the noise. However, the most
important potential contributor and the only one that will be
discussed here is the resistive poly-gate of the input
transistor. This could, if not taken into consideration, introduce a noise equal to the resistance given by \cite{sansenchang}

\begin{equation}
R_{G} = \frac{R_{i}}{12n}
\label{eq:noiseg}
\end{equation}

where R$_{i}$ is the resistance of the single poly-stripe of the width
of the input transistor M1 and n is the
number of poly-strips. As seen in Eq.~(\ref{eq:noiseg}), the
effective resistance depends on the
number of strips used in the layout. Its contribution is here
minimized by dividing the width of the input transistor in n = 60 strips, giving an effective resistance of 1.5 $\Omega$. 
The strips are also connected at both ends in order to avoid increasing its resistive gate noise contribution by four \cite{sansenchang}.

\begin{figure}[hp]
\begin{center}
\begin{tabular}{ll}
\includegraphics[width=0.5\textwidth]{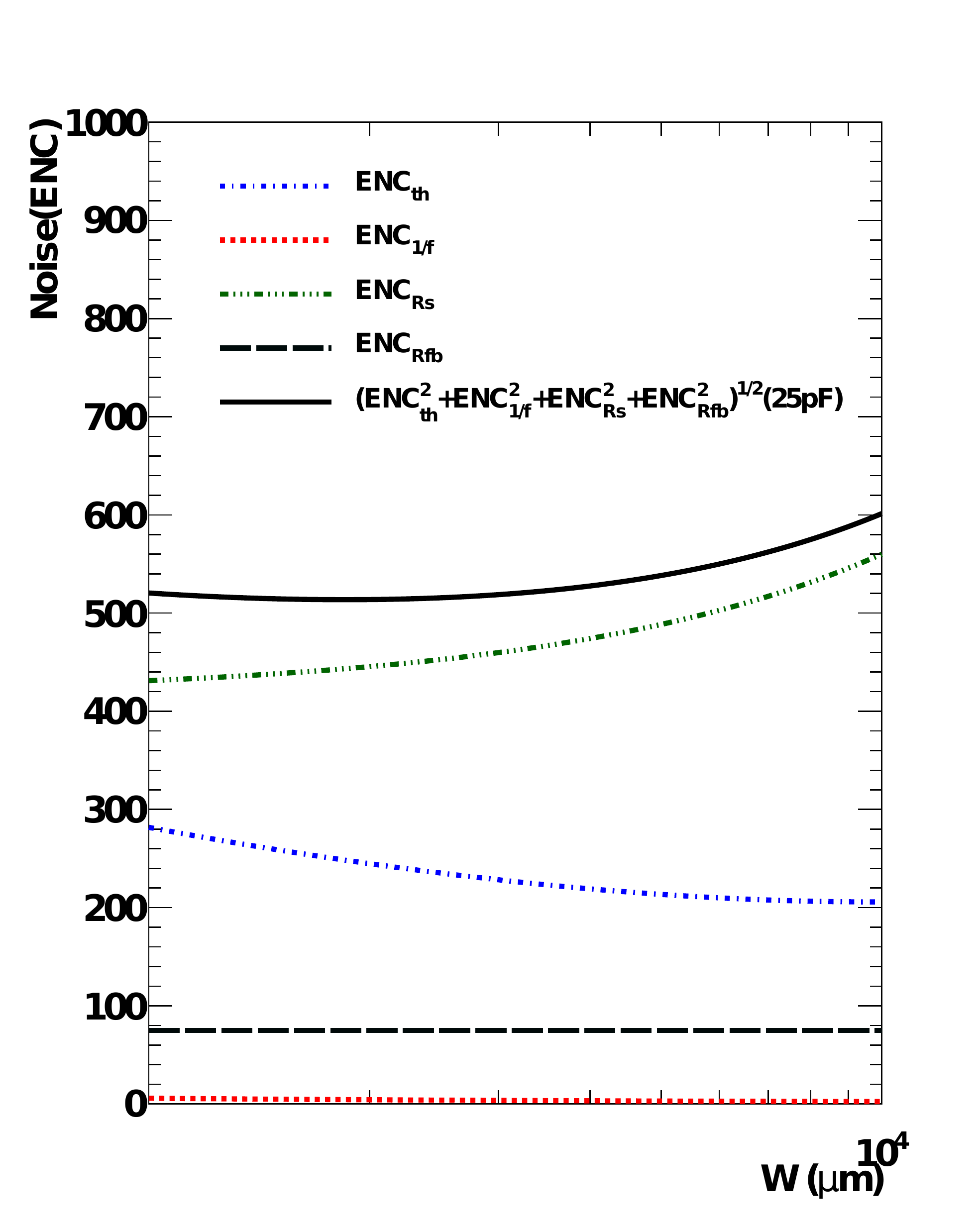}&
\includegraphics[width=0.5\textwidth]{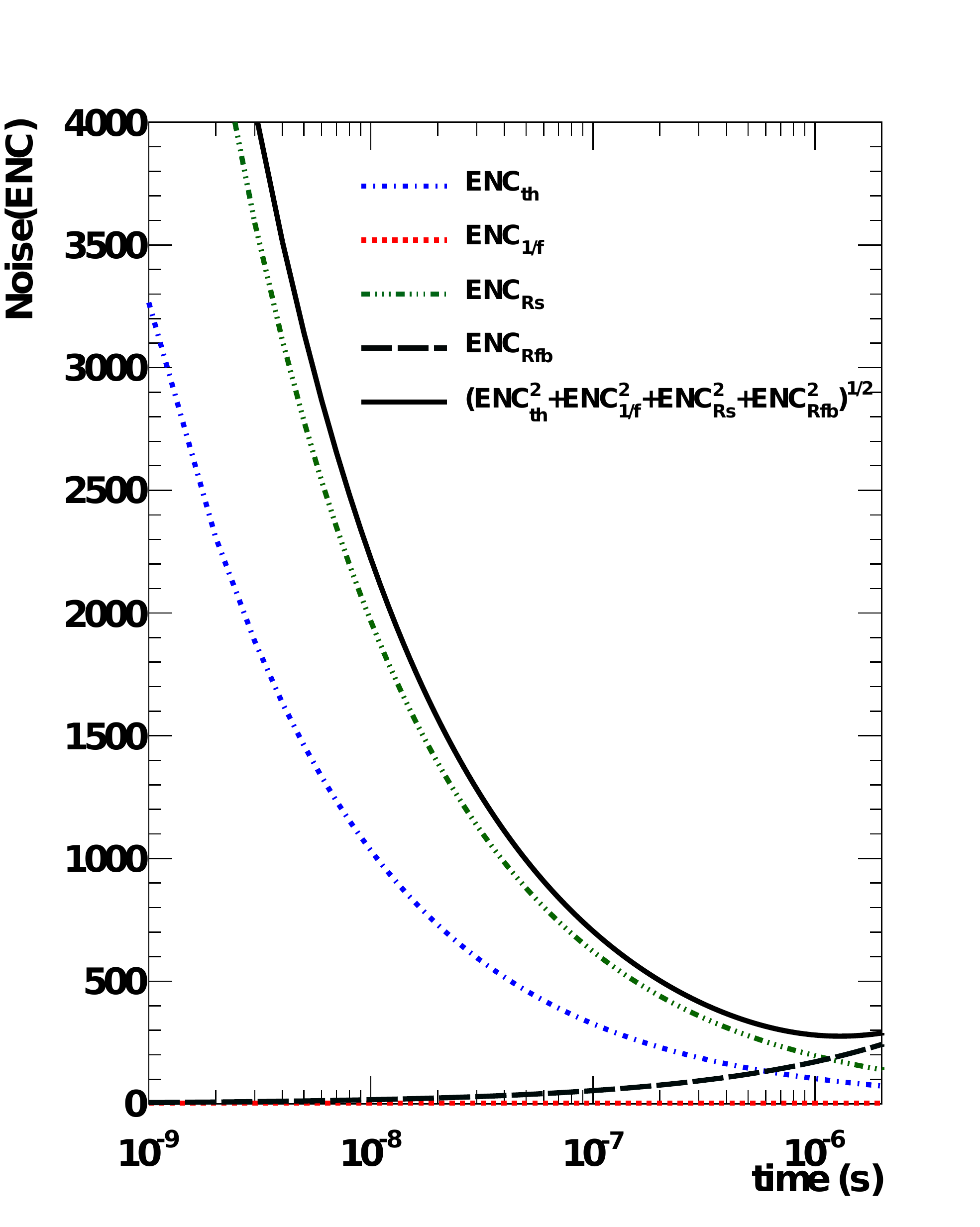}\\
\end{tabular}
\caption{The theoretical optimum noise level as function of width
  (left) and as a function of the peaking time (right). The four most important sources are quantified individually.}
\label{fig:theonoisew}
\end{center}
\end{figure}

For the given peaking time, one
can see that the ENC is proportional to the total input capacitance C$_{T}$ and increases
with the square root of the total series resistance Rs. Calculations give
here a noise of 468 electrons (Fig.~\ref{fig:theonoisew}). Thus, for noise
optimization it is important to keep this resistance as low as possible. 

Eq.~(\ref{eq:rfb}) gives the thermal noise associated with the feedback transistor
M$_{f}$, where R$_{f}$ is the equivalent resistance. This noise
contribution is inverse proportional to the value of the feedback
resistance and proportional to the shaping time constant $\tau_s$. The highest  
resistance value compatible with the requirement in terms of signal pile-up was chosen. 
Calculations convert this resistance into a typical noise of 75 electrons (Fig.~\ref{fig:theonoisew}).

Concerning the input transistor M1 of the CSA,  its  noise contribution is  dominated
by two components, the thermal noise generation in the channel and the flicker
noise given by Eq.~(\ref{eq:th}) and Eq.~(\ref{eq:overf}), respectively.
Due to the peaking time (190 ns) it is anticipated here that the thermal noise
is the dominant noise source. It is seen that its thermal
noise contribution is proportional to the total input capacitance C$_{T}$ and
inverse proportional to the square root of its transconductance $gm$. 
Therefore, in order to reduce the thermal noise, the transconductance
gm of the input transistor M1 has to be increased.
A calculated thermal noise contribution from M1 of about 238 electrons is
achieved here (Fig.~\ref{fig:theonoisew}).

The second noise source in the input transistor M1 given by
Eq.~(\ref{eq:overf}), is the flicker (1/f) noise. This noise is inverse proportional to
the gate area of the input transistor, strongly dependent on the
technology process and independent on the peaking time.
The calculation shows that a flicker noise contribution of 4 electrons
(simulation shows a contribution of about 20 electrons) is reached (Fig.~\ref{fig:theonoisew}). 

Since all the considered noise sources discussed above are uncorrelated, the
total ENC$_{T}$ can be obtained by quadratically summing all the
individual contributions:

\begin{equation}
ENC_T^2 = ENC^{2}_{th}+ENC^{2}_{\frac{1}{f}}+ENC^{2}_{R_{fb}}+ENC^{2}_{R_{S}}.
\label{eq:total}
\end{equation}

Figure~\ref{fig:theonoisew} shows the ENC contribution of each noise source, as well as
the total ENC$_{T}$ (solid line) plotted as a function of the width
(left), and as function of the peaking time (right). 
From these plots it is clearly seen that the two most dominate noise sources are the series
resistance Rs from the ESD network and the thermal 
noise of the input transistor M1. The ESD protection resistance Rs alone,
increases the overall noise of the PASA from 248e@25pF to 530e@25pF. 
Obviously, the best noise performance is achieved without any
protection, but the risk of damage will be much higher.

The dependence of the different noise sources with the peaking time
was also studied (Fig~\ref{fig:theonoisew}, right).
The optimum ENC$_{T}$ (solid line) was found to have a theoretical 
minimum at about 260 electrons with a shaping time of 1000 ns, but
this parameter was fixed to 190 ns by the requirement.
 
As seen in Fig~\ref{fig:theonoisew} (left), given the total
capacitance of 25 pF and a peaking time of 190 ns, and a current Id of 1.5~mA, a value of 
2400/0.3  was chosen for W/L  of the CSA input transistor. This gives a good noise
performance and fulfills the ALICE TPC requirements.

\section{The shaper network}
\label{sec:shaper}

\begin{figure}[hp]
\begin{center}
\includegraphics[width=0.7\textwidth]{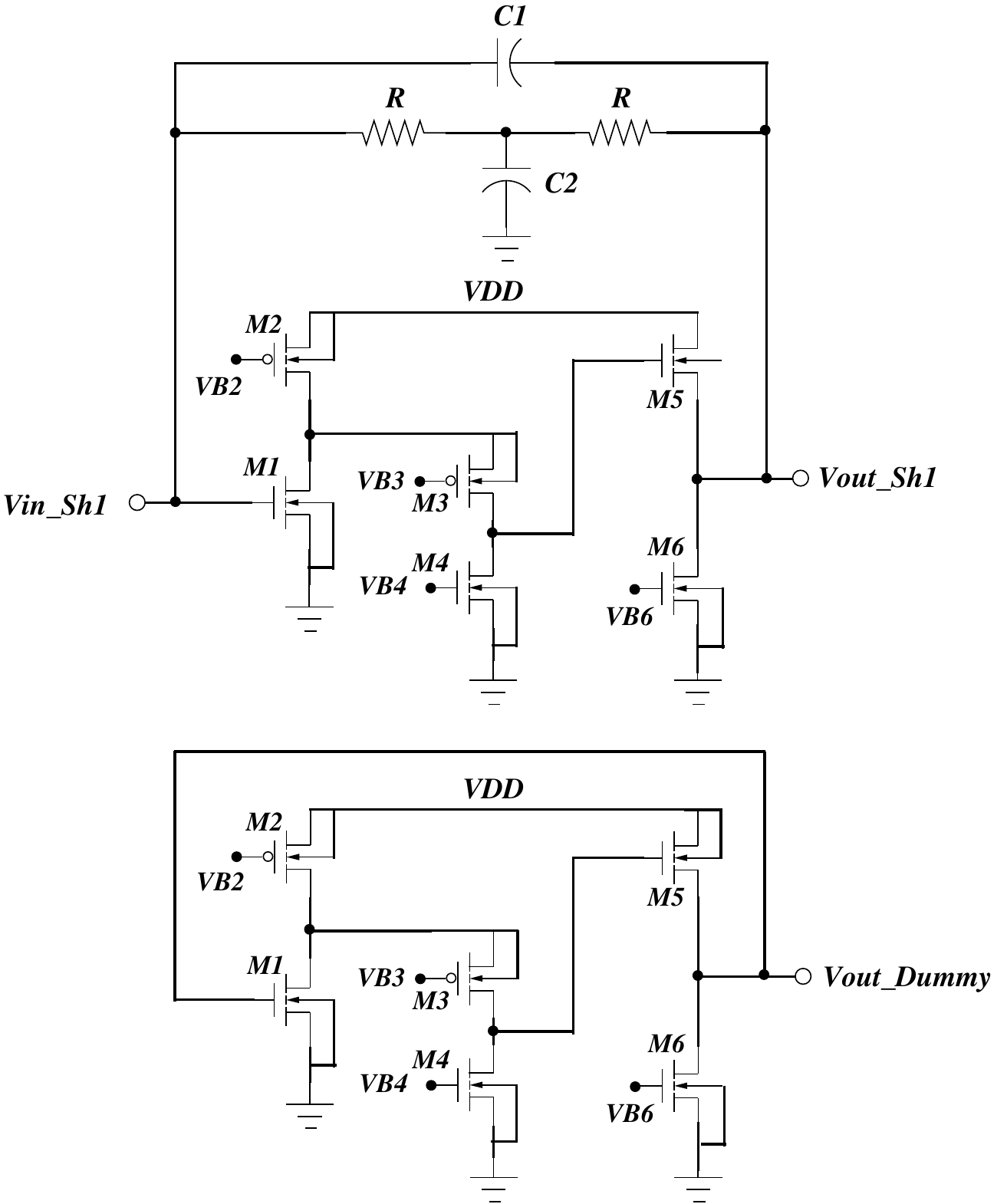}
\caption{Schematic diagram of the first shaper (upper) and the dummy amplifier (lower).}
\label{fig:1sha}
\end{center}
\end{figure}

\begin{figure}[hp]
\begin{center}
\includegraphics[width=1.1\textwidth]{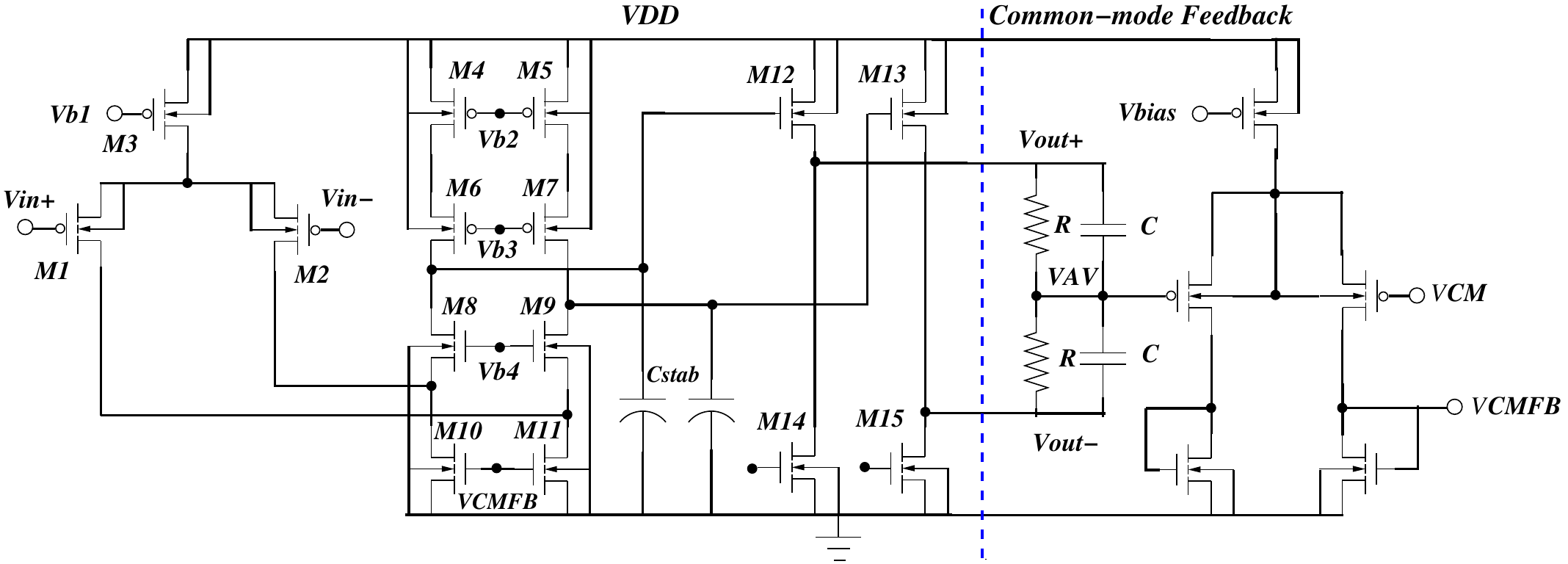}
\caption{Schematic diagram of the fully differential folded cascode circuit
  used for the second shaper (left) and the CMFB network (right).}
\label{fig:2sha}
\end{center}
\end{figure}

The function of the shaper network is to limit the bandwidth of the output signal, in order to avoid aliasing in the subsequent digitization process, and at the same time it has to optimize the overall signal-to-noise ratio.
These objectives are achieved by a semi-Gaussian shaper, which is implemented with two low-pass filter stages. Each stage consist of two second-order  bridged-T filters connected in cascade as shown in
Fig.~\ref{fig:1sha} and Fig.~\ref{fig:2sha}.

The first filter generates the first two poles and one zero in the
low-pass filter chain. The Vgs of the input transistor of the first
shaper was kept equal to the Vgs of the M1 transistor in the CSA, such that the effect of process, temperature and supply voltage variation is largely mitigated. In this context, the first shaper  (Fig.~\ref{fig:1sha}) is a  scaled-down version as the CSA.  
In order to provide the second shaper with a differential mode input, and to
track the DC level variations of the first shaper caused by the
unavoidable variations of process parameters and operating conditions, a copy of the first 
shaper connected in unity gain configuration was implemented.
This part has a combined power consumption of 1.43~mW and an open-loop gain of 75~dB.

The second shaper (Fig.~\ref{fig:2sha}) consists of a
fully differential amplifier with a folded cascode configuration and a
CMFB network. It has the same functionality as the first shaper, namely,
to implement two other poles and a zero. Together with the
differentiator CR-stage and the first shaper stage, it creates the CR-(RC)$^4$ semi-Gaussian filter. 

A continuous time CMFB network (Fig.~\ref{fig:2sha})  was implemented to prevent the output of the
fully differential amplifier from drifting to either of the two power
supplies. This establishes a stable
common-mode voltage VCM of 1.056~V at the output of the second shaper. 

The chosen CMFB network (Fig.~\ref{fig:2sha}) consists of a
resistor/capacitor network applied at the input of the CMFB
network. This configuration takes the average of the two outputs, Vout+ and Vout-,
and compares it with an externally given voltage VCM. Any deviation of
the average value of the two output voltages Vout+, Vout-  with respect to VCM is sensed and fed back
through VCMFB to the second shaper and
corrected for. This scheme ensures a fully balanced output over a voltage range limited by
the Common-Mode Range (CMR) of the two non-inverting amplifiers and the source follower of
the second shaper.

\section{The non-inverting stages}
\label{sec:nonin}

The purpose of this stage is to adapt the DC voltage level of the PASA 
outputs to the input DC levels of the ALTRO. 
The DC level VOUT+ and VOUT- is set to  0.56 V and 1.56 V for VOUT+ and VOUT- respectively by means of the external bias voltages VREFN (0.56~V), VREFP (1.56~V), and VCM (1.056 V).

\begin{figure}[hp]
\begin{center}
\includegraphics[width=0.5\textwidth]{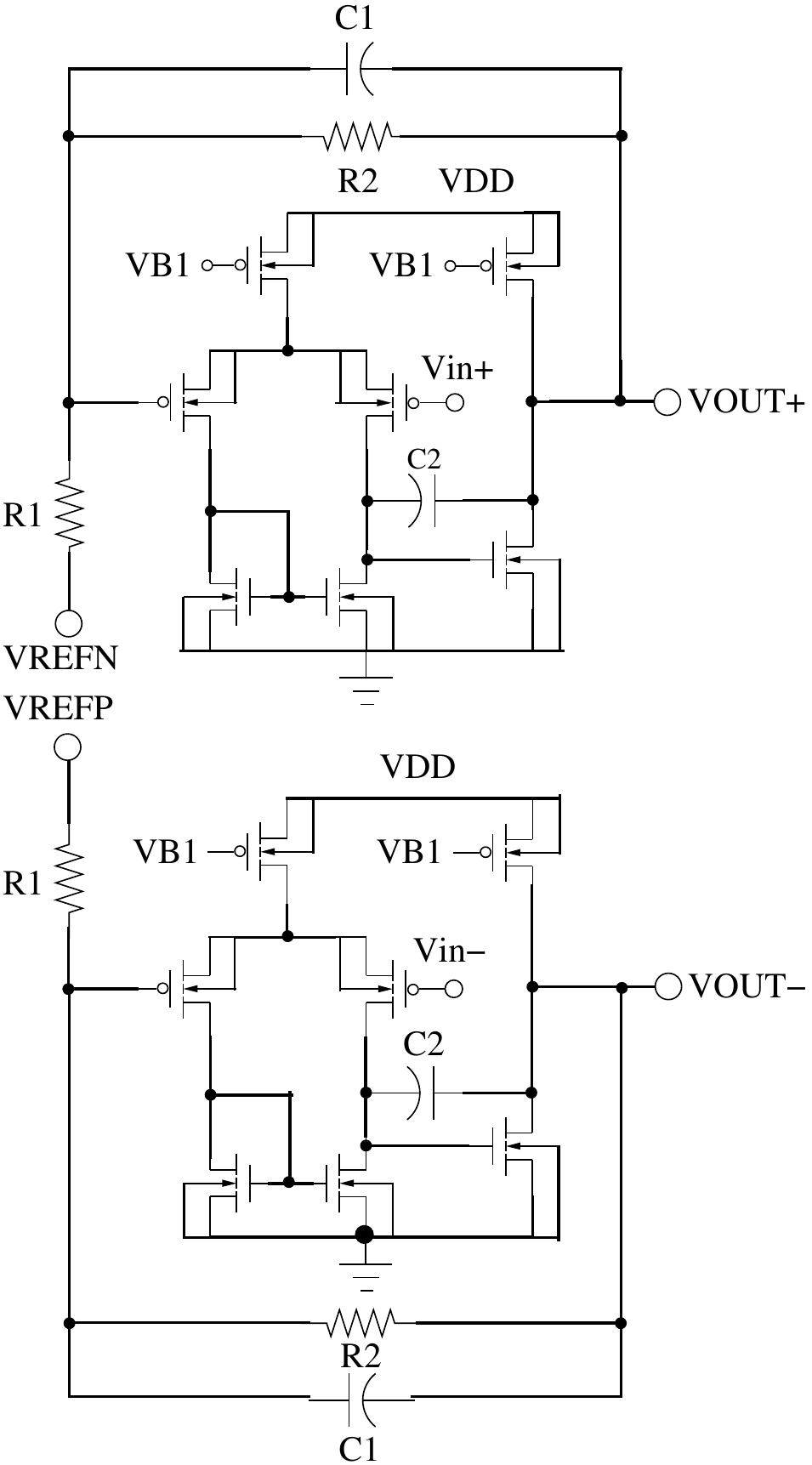}
\caption{Schematic of the two non-inverting stages.}
\label{fig:2out}
\end{center}
\end{figure}

The last stage in the PASA chain
(Fig.~\ref{fig:2out}) consists of a pseudo-differential amplifier
implemented by a parallel connection of
two equally designed Miller compensated amplifiers each with a gain of 2. 
This circuit operates in the non-inverting mode, and it uses a low
impedance reference voltage VREFP/N, to offset the output.
 
These two stages have the possibility to
change the output levels, independently from each other, according to the following equations 

\begin{equation}
VOUT+ = (1+{\frac{R2}{R1}})Vin-\frac{R2}{R1}VREFN 
\label{eq:VREFN}
\end{equation} 
and
\begin{equation}
VOUT- = (1+{\frac{R2}{R1}})Vin+\frac{R2}{R1}VREFP 
\label{eq:VREFP}
\end{equation} 
 
To prevent potential noise entering the reference path, the external reference voltages VREFP, VREFN and VCM
are internally decoupled with 120~pF each.  

\section{The threshold reference self-biased-network}
\label{sec:bias}

\begin{figure}[hp]
\begin{center}
\includegraphics[width=0.9\textwidth]{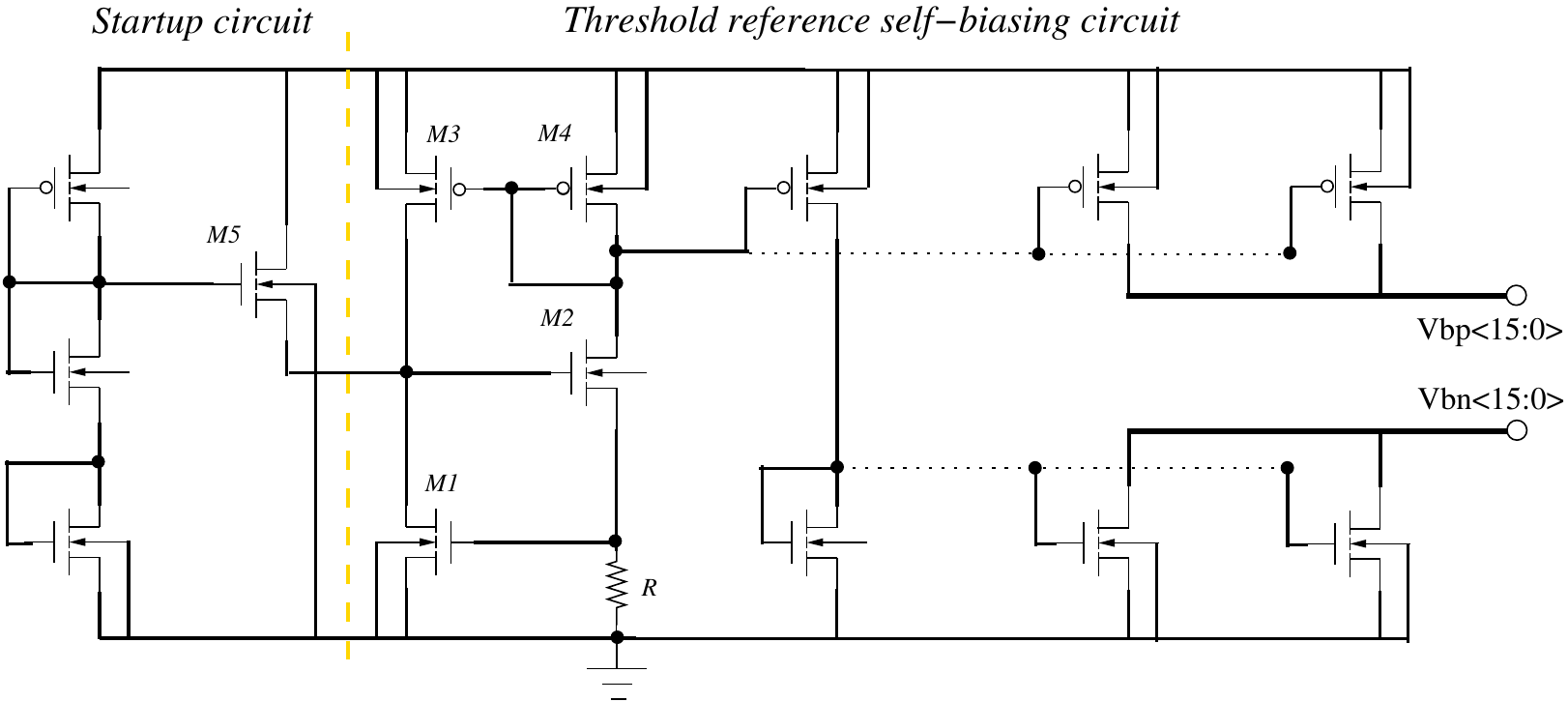}
\caption{Schematic diagram of the modified threshold reference self-biasing circuit.}
\label{fig:bias}
\end{center}
\end{figure}

All internal reference voltages of the PASA circuit are generated by a "threshold reference" self-biased network (Fig.~\ref{fig:bias}) 
is used as a bias for the PASA.
It consists of a start-up circuit, a threshold reference self-biased circuit and a chain of
current sources for both NMOS and PMOS transistors. 
The start-up circuit provides a current to the reference circuit
through M5 and, as soon as the reference circuit is
operating in the desired operation condition, the transistor M5 turns off, and it
will not draw any current under normal operation \cite{Baker}.
  
To avoid any resistive voltage drop across the width of the chip 
the distribution of the bias voltages is done by 32 reference currents
distributed over the 16-channels. The bias current is then channeled through
a diode connected transistor at each bias point to provide locally the bias
voltages for the individual transistors. 

The threshold self-biased network is placed in the center of the chip 
(see Fig.~\ref{fig:layout}).

\section{Layout and floor-planning}
\label{sec:layout}

To realize a high quality PASA it is very important to take care of the
matching and noise issues.  Hence, the common-centroid layout technique was used for parallel and equally
sized transistors to avoid errors caused by the gradient effects across
the chip, such as temperature, stress and gate-oxide thickness.
To minimize the possible mismatch induced by etch undercutting during
fabrication, dummy poly strips were used for resistors, capacitors and transistors.
As mentioned in subsection~\ref{subsec:theo}, a finger structure is
used for the input transistor M1 to reduce the poly gate resistive
noise. This structure also has lower C$_{bs}$ (bulk-source capacitance) and C$_{bd}$
(bulk-drain) capacitances and it is convenient for laying out a MOSFET with large W/L
ratio ~\cite{sansenchang}.

\begin{figure}
\begin{center}

\hspace*{-1.65cm}\includegraphics[width=1.3\textwidth]{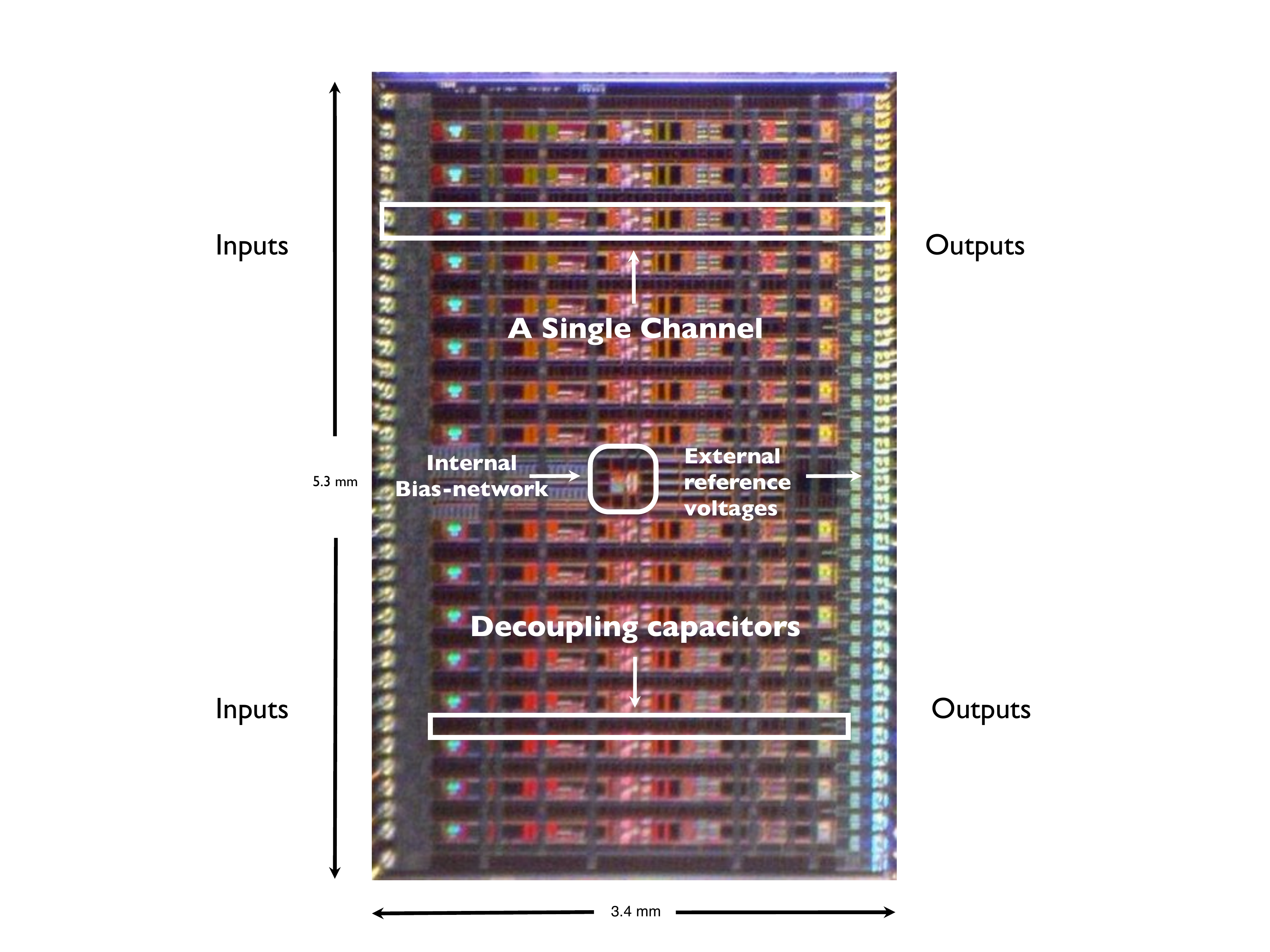}
\caption{A picture of the produced ALICE TPC PASA.}
\label{fig:layout}
\end{center}
\end{figure}

Since noise can couple into the bias nodes and in order to avoid that large
signals parasitically couple into the more sensitive analog parts, all
bias voltages, especially at the CSA part, are heavily
decoupled/filtered to reduce the amplitude of the noise and to keep the
internal cross-talk to a minimum. 

After the main layout was finished, all free space was filled with substrate
contacts,  to both the substrate and the
wells. 

A picture of the PASA layout is shown in Fig.~\ref{fig:layout}.
It has 16 input, 32 output, 16 GND pins, 8 VDD
pins and 3 input pins for the reference voltages. The input and
output pins are distributed along the length of the chip. The 16-channels, divided in two groups of 8 channels are placed on
each side of the threshold self-biased bias network, seen at the center of the chip.
The decoupling capacitors together with pads connected to ground are
placed adjacent to each channel, creating a physical distance between the channels
that helps reducing the cross-talk. In addition each channel is surrounded by a guard ring
connected to the substrate which isolates them from each other and
further helps to reduce cross-talk.
 
The final layout of the TPC PASA has a width of 5.3~mm and a length of
3.4~mm,  giving a total  area of 18~mm$^2$. 

\section{Functional analysis prior to submission using simulation}
\label{sec:simu}

\begin{figure}[p]
\begin{center}
\hspace*{-0.5cm}\includegraphics[width=0.5\textwidth,angle=-90]{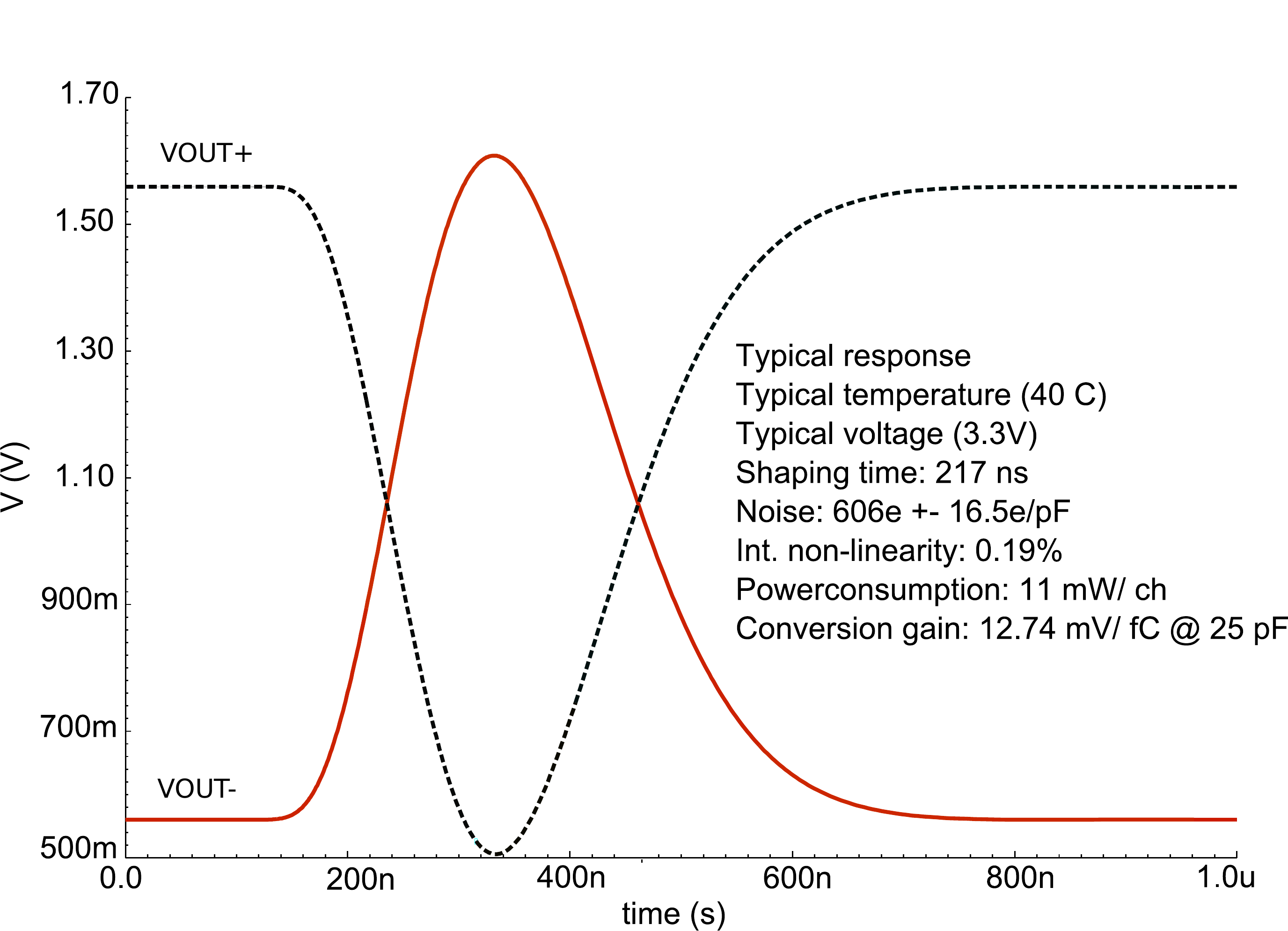}
\end{center}
\caption{A typical impulse response of the PASA for an input charge of
  165fC. The curves for VOUT+ (dashed) and
VOUT-  (solid) are shown. The baseline voltage of VOUT+  and VOUT- are 1560~mV and 560~mV,
respectively.}
\label{fig:simpulse}
\end{figure}

\begin{figure}[p]
\begin{center}
\includegraphics[width=0.5\textwidth,angle=-90]{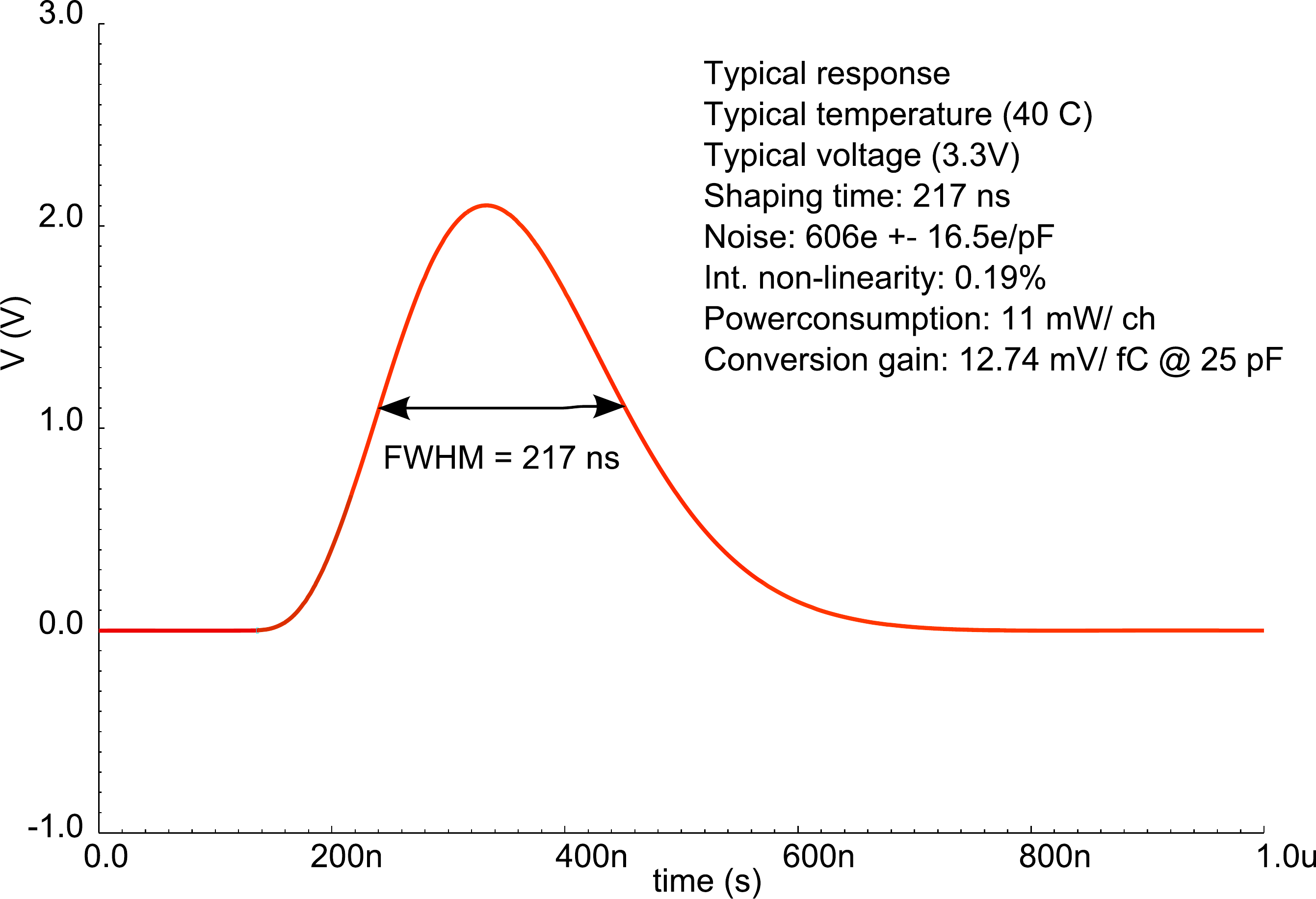}
\caption{A typical simulated differential pulse response of the PASA for an input charge of
  165~fC.}
\label{fig:simoutput}
\end{center}
\end{figure}

In this section the result of the most relevant simulations are illustrated. They all refer to simulations performed on the circuit back-annotated with the parasitic capacitances extracted from the circuit layout, with an input capacitive load of 25 pF and an input charge of 165 fC. In the design optimization phase, special emphasis was put on  understanding the circuit  behavior for different process parameter, temperature and voltage changes. 
The simulated typical PASA impulse response is shown 
in Fig.~\ref{fig:simpulse} and Fig.~\ref{fig:simoutput},  which illustrate each of the two output signal polarities (VOUT+ and VOUT-) and the differential one respectively.
The simulations show the following circuit features: a peaking time of 190 ns, a full width half maximum (FWHM)
of 217 ns, a conversion gain of 12.7 mV/fC for the combined outputs, a return to the signal baseline within 550 ns and no undershoot. Under the same simulation conditions the circuit features an equivalent noise charge (ENC) of about 600 electrons. 
 
It shout be noticed that the total capacitance at the circuit input influences the noise, the peaking time and the conversion gain.

\begin{figure}[p]
\begin{center}
\includegraphics[width=0.8\textwidth]{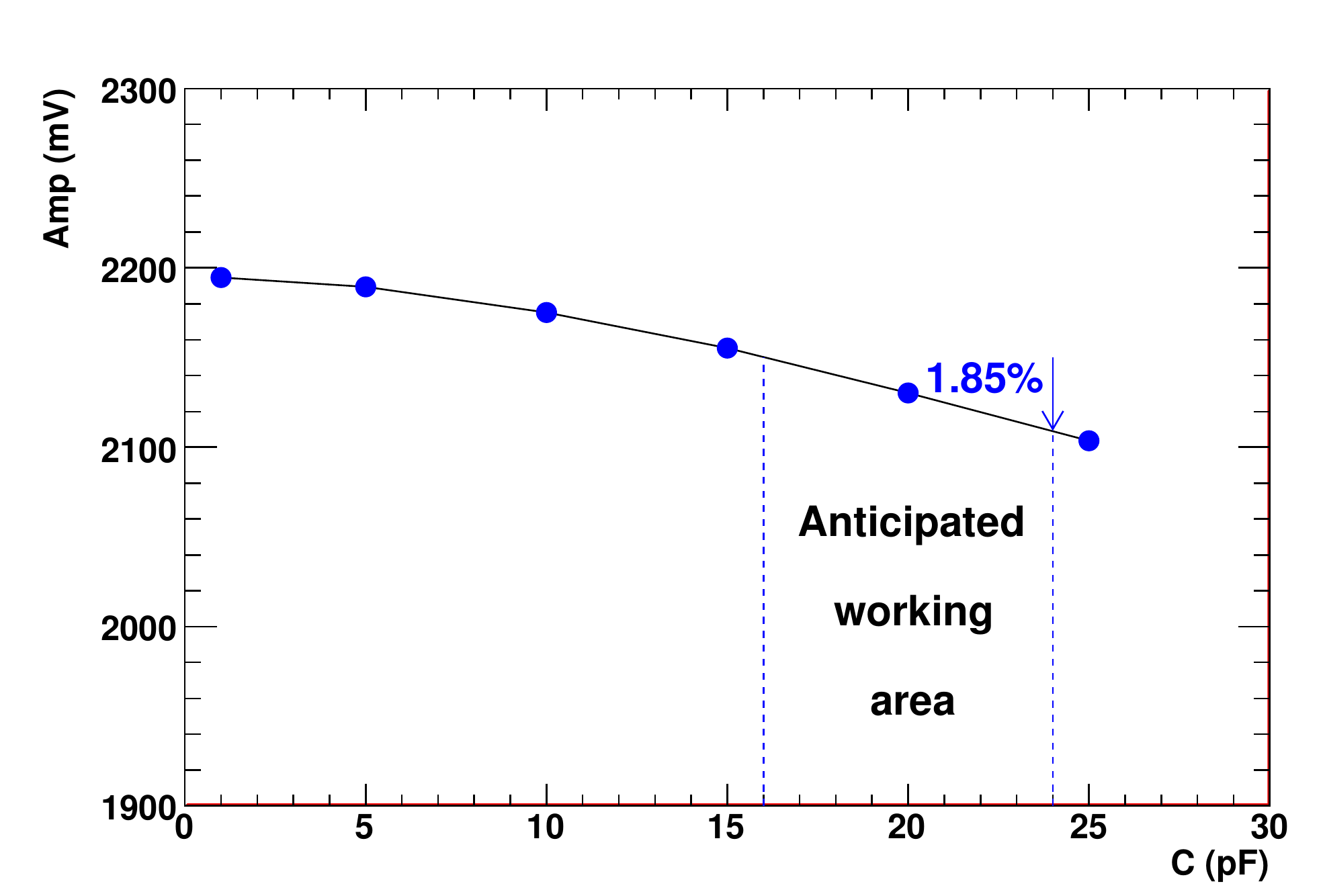}
\caption{Amplitude variation of the PASA impulse response as a function
  of total detector capacitance.}
\label{fig:ampcap}
\end{center}
\end{figure}

In the case of the ALICE TPC the capacitance at the PASA input can vary from a minimum of about 10 pF to a maximum of about 25 pF. This includes the contribution from the pad plane and the flexible cable that connects the pad plane to the front-end card (1-20 pF), PCB traces on the front-end card (1-3 pF), and from the package capacitance of about 2 pF. This capacitive spread from 10 to 25pF converts into a variation of the conversion gain from 13.3 mV/fC to 12.74 mV/fC and peaking time variation from 181 ns to 190 ns, is due to the reduction of the bandwidth of the CSA.
The pulse amplitude as a function of the detector capacitance has been
evaluated and is shown in Figure~\ref{fig:ampcap}.

\begin{figure}[p]
\begin{center}
\includegraphics[width=0.6\textwidth,angle=-90]{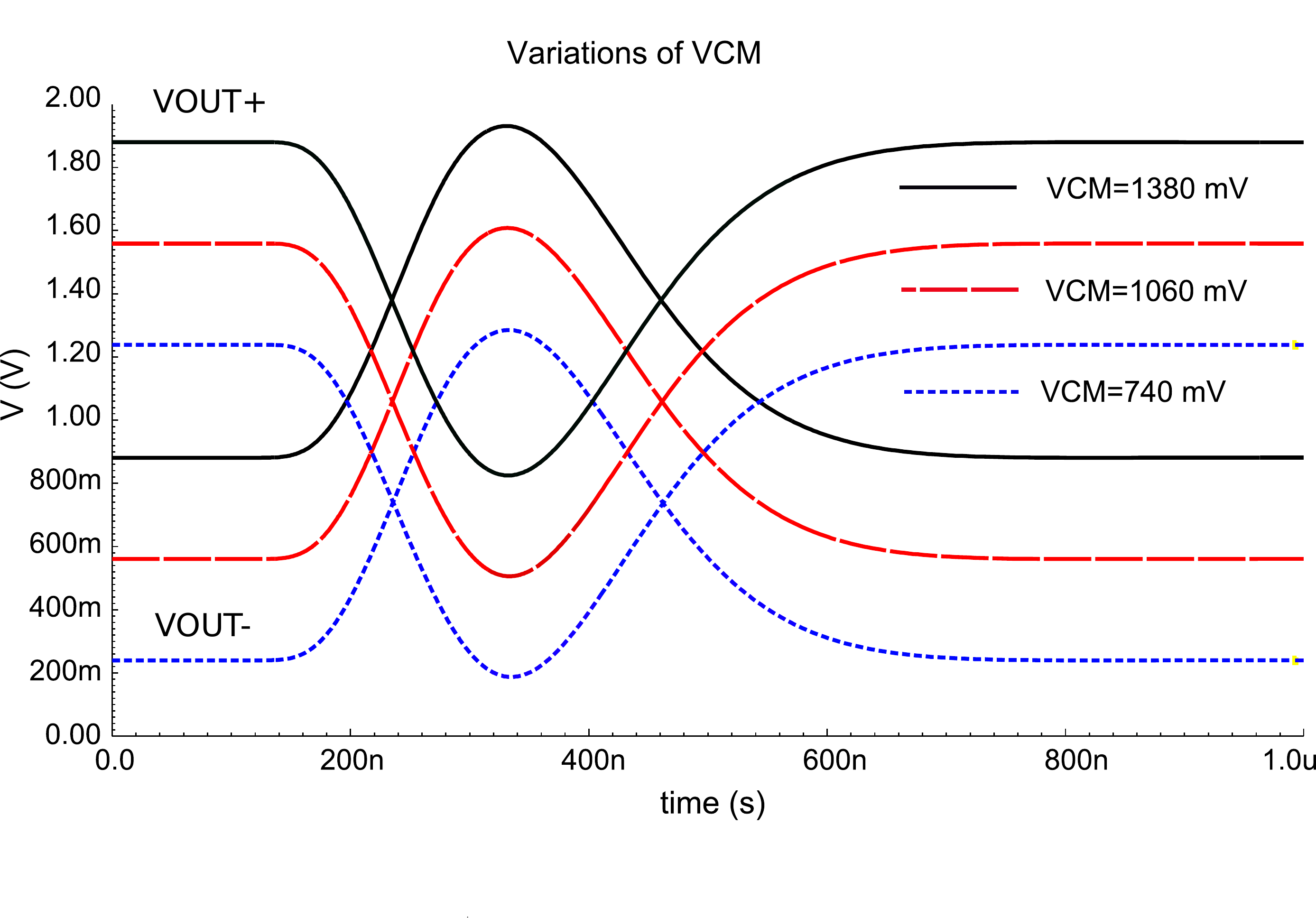}
\caption{DC output level variation of VOUT+ and VOUT- due to variation  of VCM. }
\label{fig:simlevelVCM}
\end{center}
\end{figure}

\begin{figure}[p]
\begin{center}
\includegraphics[width=0.9\textwidth,angle=-0]{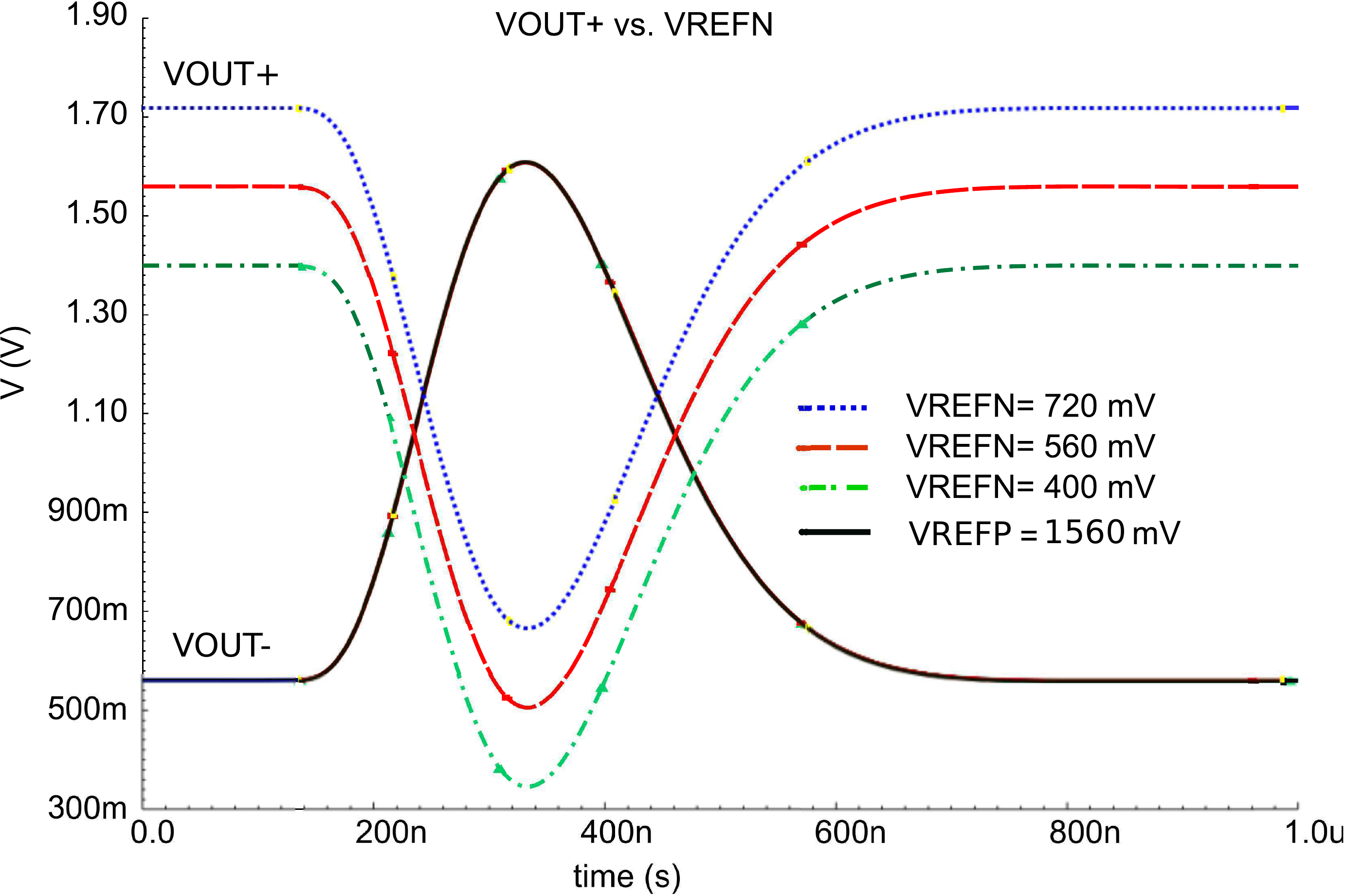}
\caption{DC output level variation of VOUT+ due to variation of VREFN. }
\label{fig:simlevelVREFN}
\end{center}
\end{figure}

Fig.~\ref{fig:simlevelVCM} shows the signal response of the
common-mode voltage VCM when changed externally +/- 320 mV around 1.060 V.
No deviation of the signal is seen.

Fig.~\ref{fig:simlevelVREFN} shows the signal response  and DC response
of VOUT+ as a function of VREFN. A shift VREFN from 400 mV to 760mV was implemented. 
The circuit responded as given by Eq.~\ref{eq:VREFN} and
Eq.~\ref{eq:VREFP}. Also here, no deviation from typical response is observed.  

\begin{figure}[p]
\begin{center}
\includegraphics[width=0.9\textwidth,angle=-0]{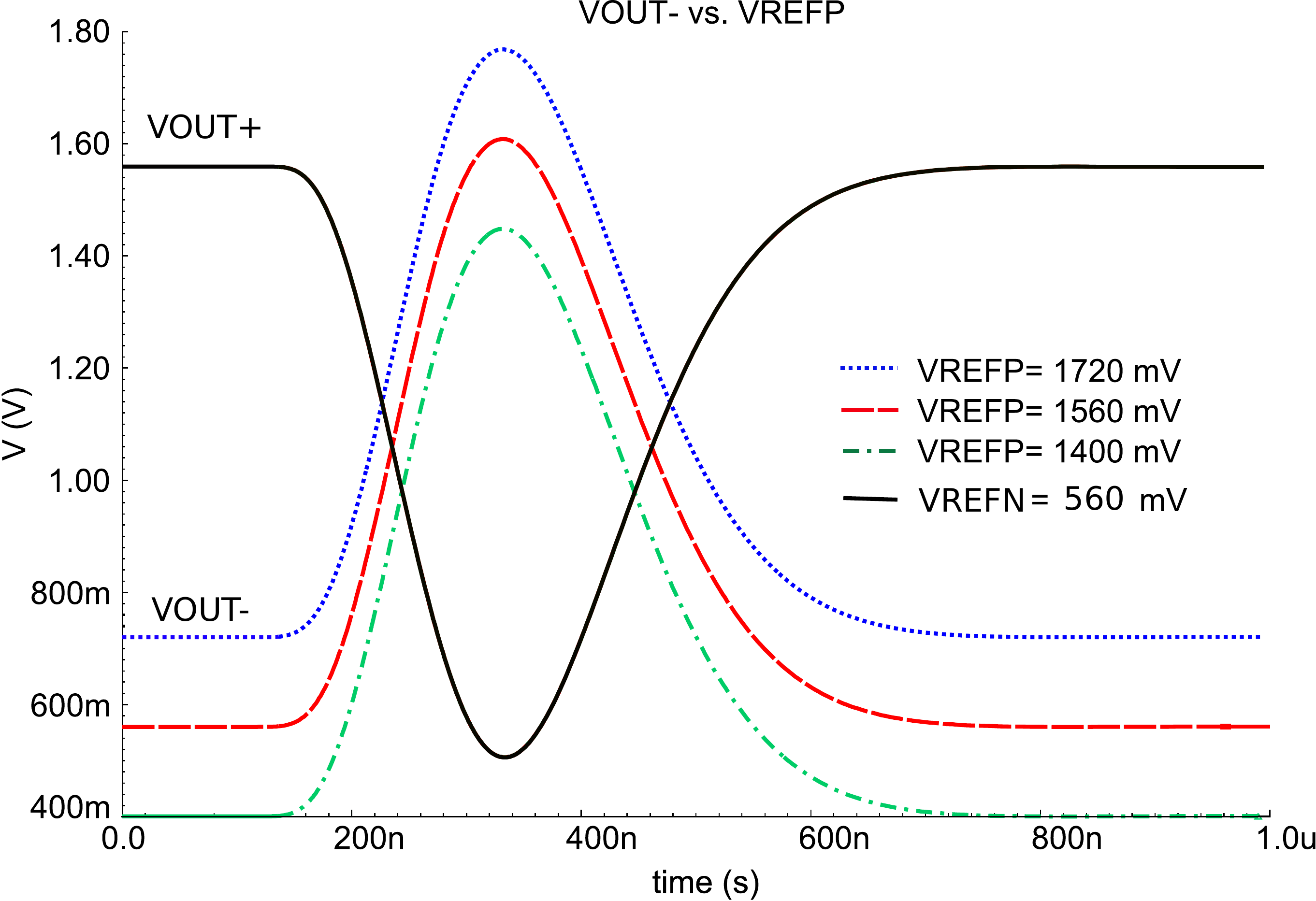}
\caption{DC output level variation of VOUT- due to variation of VREFP. }
\label{fig:simlevelVREFP}
\end{center}
\end{figure}

Fig.~\ref{fig:simlevelVREFP} shows the signal response of the
common-mode voltage when changed externally +/- 320 mV around 1.060 V.
No deviation of the signal is seen.

The DC levels at the output of the PASA, VOUT+ and VOUT-, are 
controlled by three externally given reference voltages (VREFP, VREFN and VCM).
The response at the output of the PASA when VCM is varied between 740~mV and 1380~mV
is shown in Fig.~\ref{fig:simlevelVCM}. Except for the expected common-mode change,
no change in conversion gain and peaking time was seen.

The variation of the manufacturing process parameters causes the variation of the circuit electrical parameters, e.g. the values of transistor threshold, resistor and capacitors. Therefore, one of the challenges in analog chip design is to ensure that the unavoidable variations of the circuit electrical parameters do not lead to an excessive variation of its functional parameters (e.g. conversion gain, peaking time, power consumption and noise). 

\begin{figure}[p]
\begin{center}
\includegraphics[width=0.6\textwidth,angle=-90]{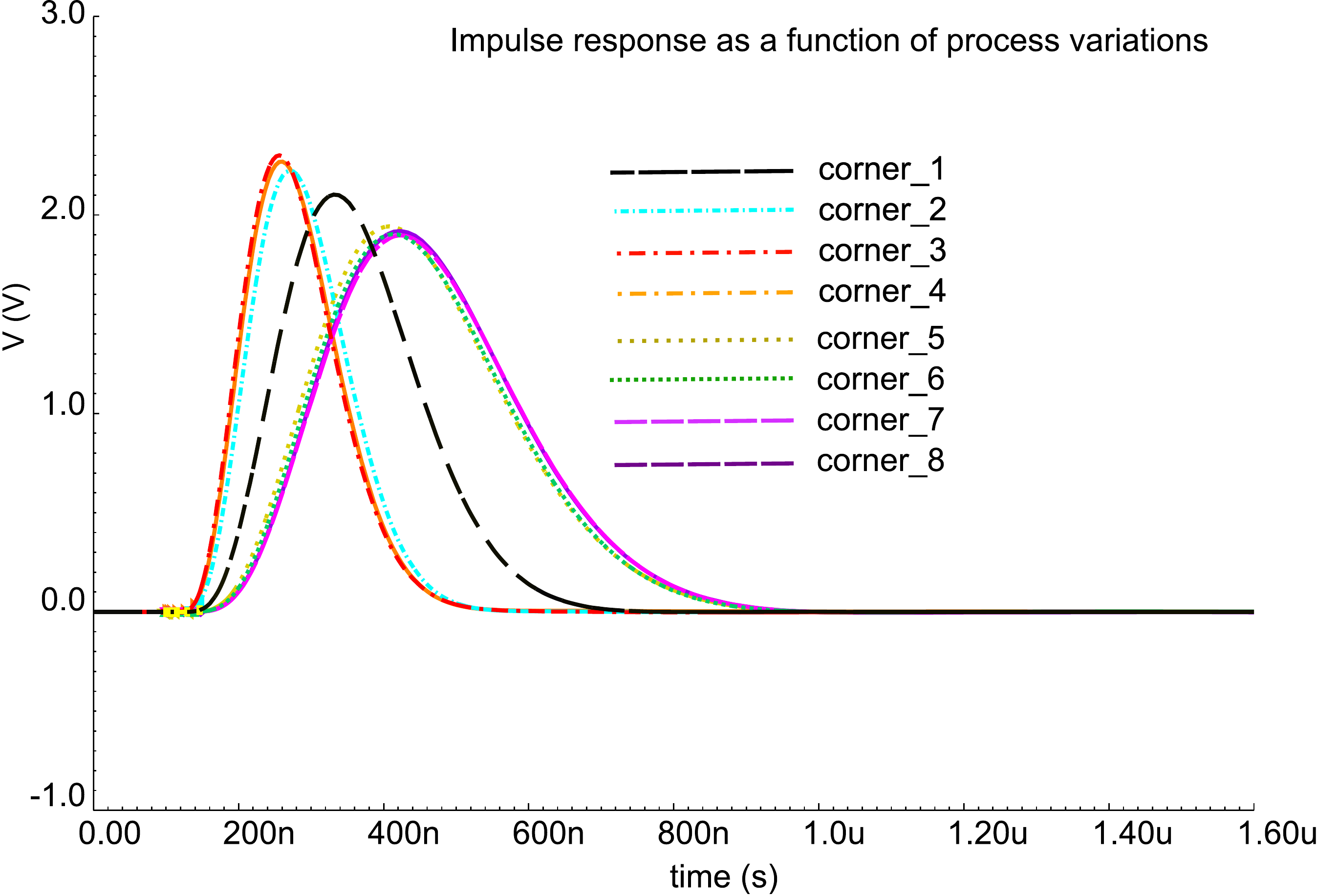}
\caption{PASA impulse response for the eight recommended corner combinations by AMS.}
\label{fig:corners}
\end{center}
\end{figure}

How this has been achieved in the design of the ALICE TPC PASA is described in section 7. The simulation results for the
worst case variation of the foundry manufacturing process parameters is  shown in Fig.~\ref{fig:corners} and (Table~\ref{tab:req1}).

\begin{table}[htb]
\begin{center}
\begin{tabular}{|c||c|c|}\hline
    Parameter          & Specific.  &Simulated     \\ \hline \hline
    Noise              &$<$ 1000e   &$\approx$ 600e(25pF)  \\ \hline
    Shaping time [ns]  & 190        & 217   \\ \hline
    Non-linearity      &$<$ 1\%     & 0.19\%   \\ \hline
    Crosstalk          &$<$ 0.3\%   & -          \\ \hline
    Baseline variation & -          & -   \\ \hline
    Conv. gain [mV/fC] & 12         & 12.7@25pF \\ \hline
    Powercon.[mW/ch]   & $<$ 20     & 11         \\ \hline
\end{tabular}
\end{center}
\caption{Compilation of the simulation results of the main parameters compared to the specifications.}
\label{tab:req1}
\end{table}

\clearpage
\section{Measurement results and system behavior}
\label{sec:meassys}

In order to qualify the ALICE TPC PASA the necessary test equipment was
developed and built at the Technische Universit\"at Darmstadt
(TU-Darmstadt, Germany) and at the Lund University (Sweden).
A comprehensive description of the test setup can be found in \cite{testsetup}; only a short explanation of the
test procedure will be given here. 

The functionality of the chip is characterized by injecting a charge generated by a 14-bit
DAC placed on the test board into the 16 PASA inputs, one at the time. The full set of tests for all 32 outputs signals were performed for 3 values of the supply
voltages (3.0,  3.3 and 3.6 V). The PASA output signals were then converted by a 12-bit
40-MSPS ADC. From these measurements the following values were extracted:  
the power consumption, the conversion gain, the peaking
time, the noise and the linearity.
Initially the static power
consumption was measured. If the power test fulfills the acceptable power
consumption region, the test continue, if not, the test was terminated and the chip was
characterized as non-functional.

Since the ALICE TPC PASA was adopted for the upgrade of the STAR TPC
at the Brookhaven National Laboratory (BNL), and went through
exactly the same test procedure, the results
obtained for both the 47637 ALICE TPC PASA chips and for the 22795 chips from the
STAR TPC production are presented here.
 
In subsection~\ref{subsec:sys} the system behavior of the
ALICE TPC PASA is shown.

\subsection{Measurement results}

\begin{figure}[htb]
\begin{center}
  \includegraphics[width=0.9\textwidth]{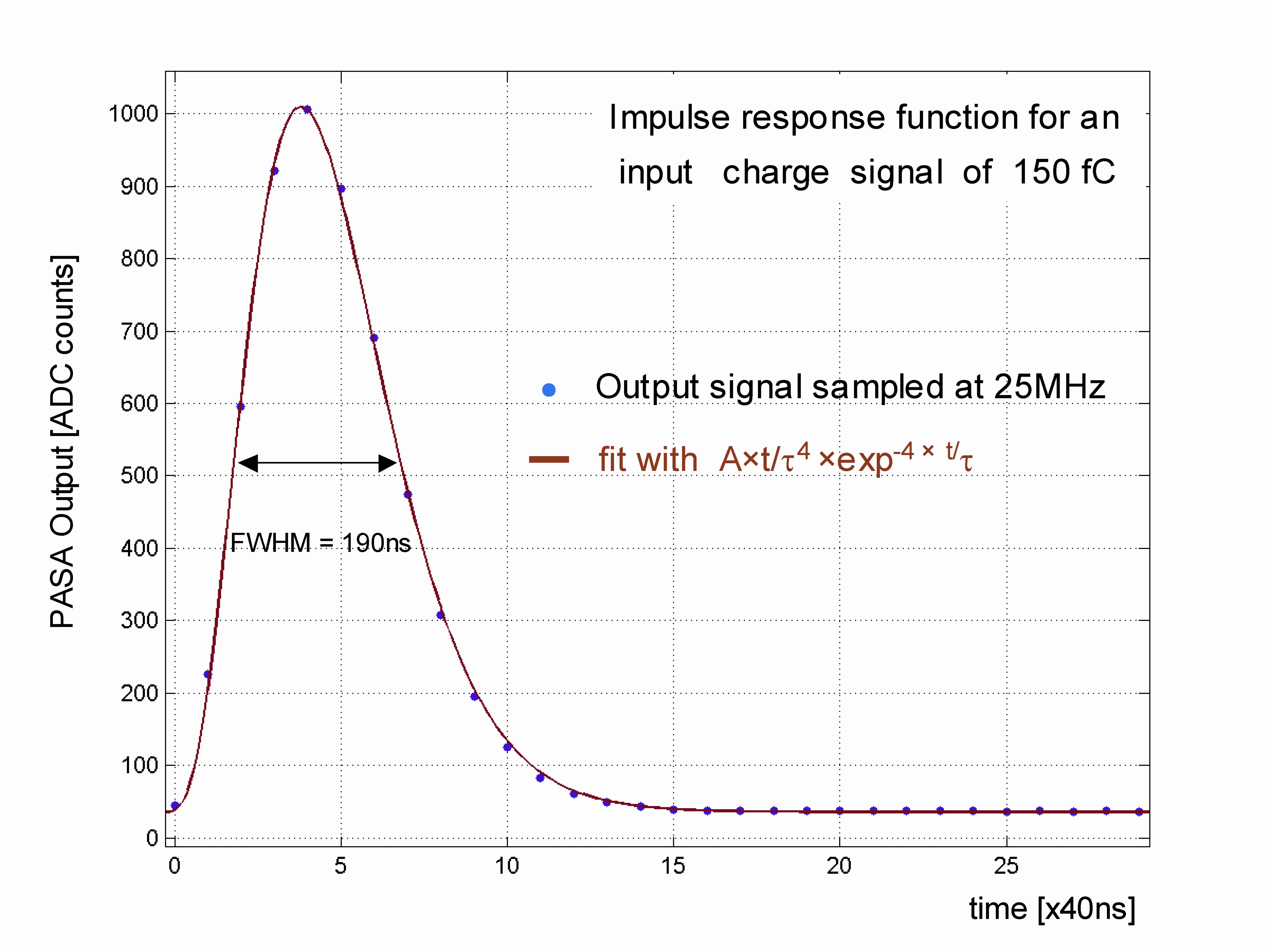}
\caption{Typical measured response of the ALICE TPC readout
  electronic. The plot shows a measured output response with an ideal
  gamma-4 response function. }
\label{fig:bild6}
\end{center}
\end{figure}

Figure \ref{fig:bild6} shows the first measured output pulse shape on one engineering
sample of the ALICE TPC production. The measurement was done on a fully equipped
front-end board for a single channel of a randomly chosen chip.
For an input charge of 150 fC the conversion gain is
12.7~mV/fC, the peaking time is 160 ns and the FWHM is 190 ns. This pulse is fitted with the ideal gamma-4 response function and
shows an almost perfect matching with the measured waveform. 
The channel integral non-linearity was found to be 0.2\% over the full dynamic range
of 150~fC, less than the 1\% given by the specifications. 
The single channel had a noise value below 570 e (r.m.s.) for an
input capacitance of 12 pF, and a channel-to-channel cross-talk was below
-60 dB.

The general performance of the full custom PASA chip is listed in
Tab.~\ref{tab:tested}, and the distribution of the key parameters are 

\begin{figure}[p]
\begin{center}
\hspace*{-0.9cm}\includegraphics[width=1\textwidth]{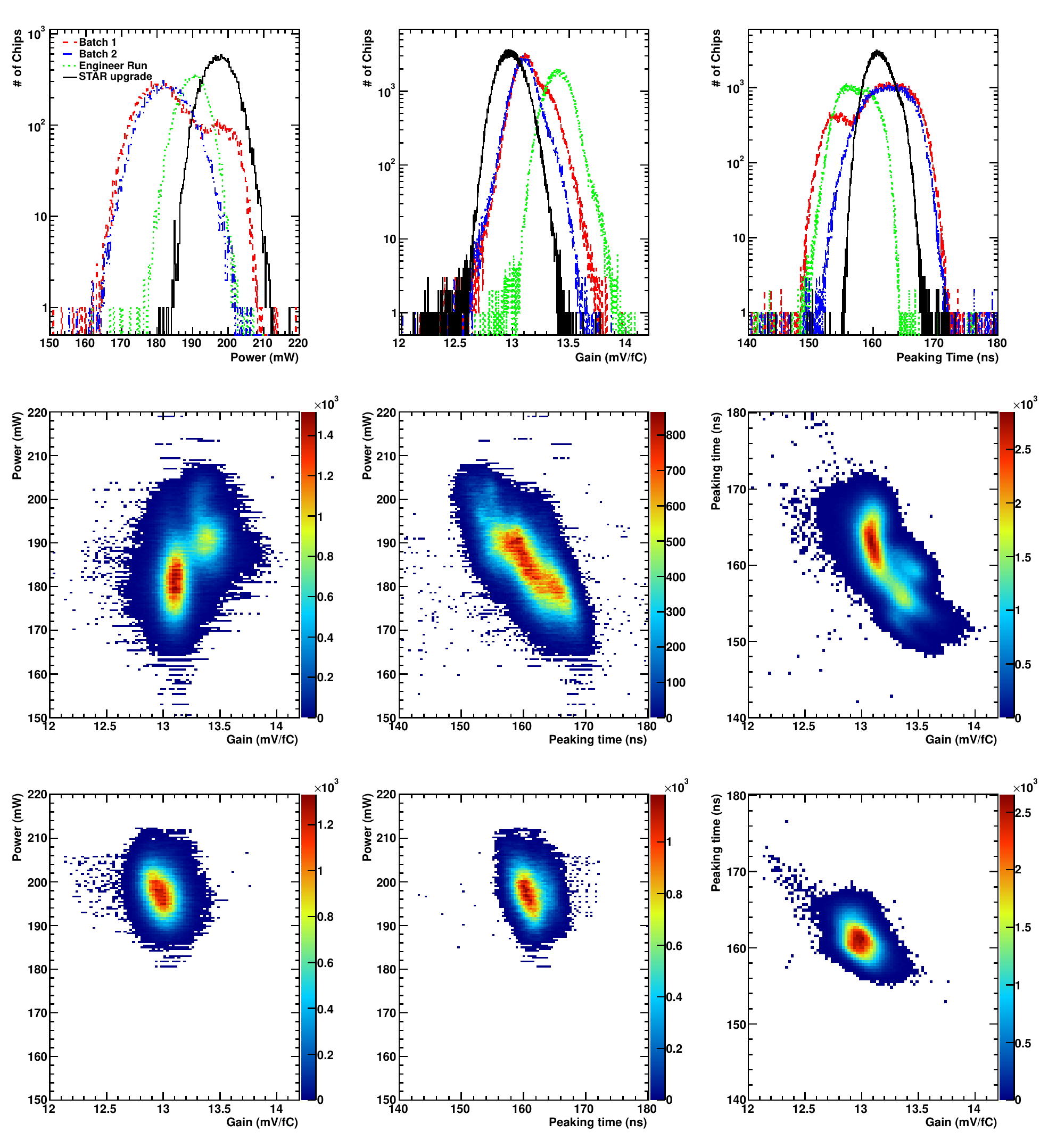}
\caption{Distribution of the power consumption (top left), conversion
  gain (top right) and peaking time (top right) for the engineering run (dotted), the two batches
  from the ALICE TPC PASA production (dashed and dashed-dotted),  and the total STAR TPC chip upgrade
  production (solid). Middle and bottom rows show from left to right the power versus conversion
  gain distribution, the power versus peaking time and the peaking time versus
  conversion gain for the ALICE TPC PASA and for the STAR upgrade
  production, respectively.}
\label{fig:testall}
\end{center}
\end{figure}

\begin{table}[h]
\begin{center}
\begin{tabular}{|c|c|c|c|}\hline
    Parameter          & Specific.       &Simulated            &Tested             \\ \hline\hline
    Noise              &$<$ 1000e        &385e(12 pF)          & $\approx$ 385e(12pF)  \\ \hline
    Shaping time [ns]  & 190             & 212(12 pF)          & $\approx$ 190     \\ \hline
    Non-linearity      &$<$ 1\%          & 0.19\%              & 0.2\%            \\ \hline
    Crosstalk          &$<$ 0.3\%        & -                   & $<$0.1\%   \\ \hline
    Baseline variation & -               & -                   & +-80mV             \\ \hline
    Conv. gain [mV/fC] & 12            & 12.74@25pF            & $\approx$ 12.8@12pF    \\ \hline
    Powercon.[mW/ch]   & $<$ 20          & 11                  & 11.67              \\ \hline
\end{tabular}
\caption{Summary of the simulated and tested values of the ALICE TPC PASA.}
\label{tab:tested}
\end{center}
\end{table}

presented in Figure ~\ref{fig:testall}.
For the ALICE TPC PASA an acceptability region of $\pm$ 20\% from the mean power
consumption, in this case of about 187~mW, was defined.
A total of 307 chips  was found to have a power consumption outside
the acceptance region. Of these, 180 chips had a power
consumption outside the maximum measurable power consumption of
285~mW, and then characterized as non-functional. 
The small difference in the average power consumption between the
distributions is due to the unavoidable variation in the process parameters from
different production lots as discussed in section 8.

Conversion gain mean values of 13.4
mV/fC@5 pF, of 13.2 mV/fC@5 pF and of 13 mV/fC@5 pF were found
for the engineering run, for the production run and for the STAR upgrade production, respectively, as seen in Figure~\ref{fig:testall}.
A small difference in the conversion gain is also seen, as is expected from chips coming from different lots or productions.

The measured peaking time (Fig. \ref{fig:testall}, top
right) is typically around 160~ns, a bit shorter than the expected value from the simulation
results. This is a known artefact seen in several
Multi-Project Wafer (MPW) runs. Therefore, to cope with this, the circuit was
designed with an slightly larger shaping time equal to the deviation
experienced from these previous MPW submissions.

Taking into consideration the known deviation in peaking time between
simulation and measurement, the measurement results show very good
agreement between simulation and measurement. An MPW production of
a new chip with similar topology and design strategy done in 2005
~\cite{4} does not show this shift in peaking time.

The bi-dimensional plot of the conversion gain versus the power consumption,
the peaking time versus the power consumption and the conversion gain versus the
peaking time are shown in Fig.~\ref{fig:testall} also. 
Relatively good distributions of the conversion gain versus power
consumption are seen for the ALICE TPC PASA and for the STAR upgrade productions, and most of the channels show a
conversion gain between 12 mV/fC and 14 mV/fC and as expected, it does not show any visible correlation with the power consumption. 
Since the STAR upgrade production consists of a single lot of 25 wafers, it shows a narrower distribution than that for the ALICE production, which consist of two lots of 25 wafers. The same is also observed for the
individual lots for the ALICE TPC PASA productions. Very few chips are outside the acceptance region.

The peaking-time versus the power consumption shows a weak
correlation for the ALICE TPC PASA and no correlation for the STAR
upgrade production. The reason is clearly the shift in process
parameters from different productions in the ALICE TPC PASA.

The conversion gain versus peaking time shows similar behavior as the
previous distribution also due to different productions. Very few chips are outside the acceptance region.

\begin{figure}[p]
\begin{center}
\hspace*{-0.9cm}\includegraphics[width=1\textwidth]{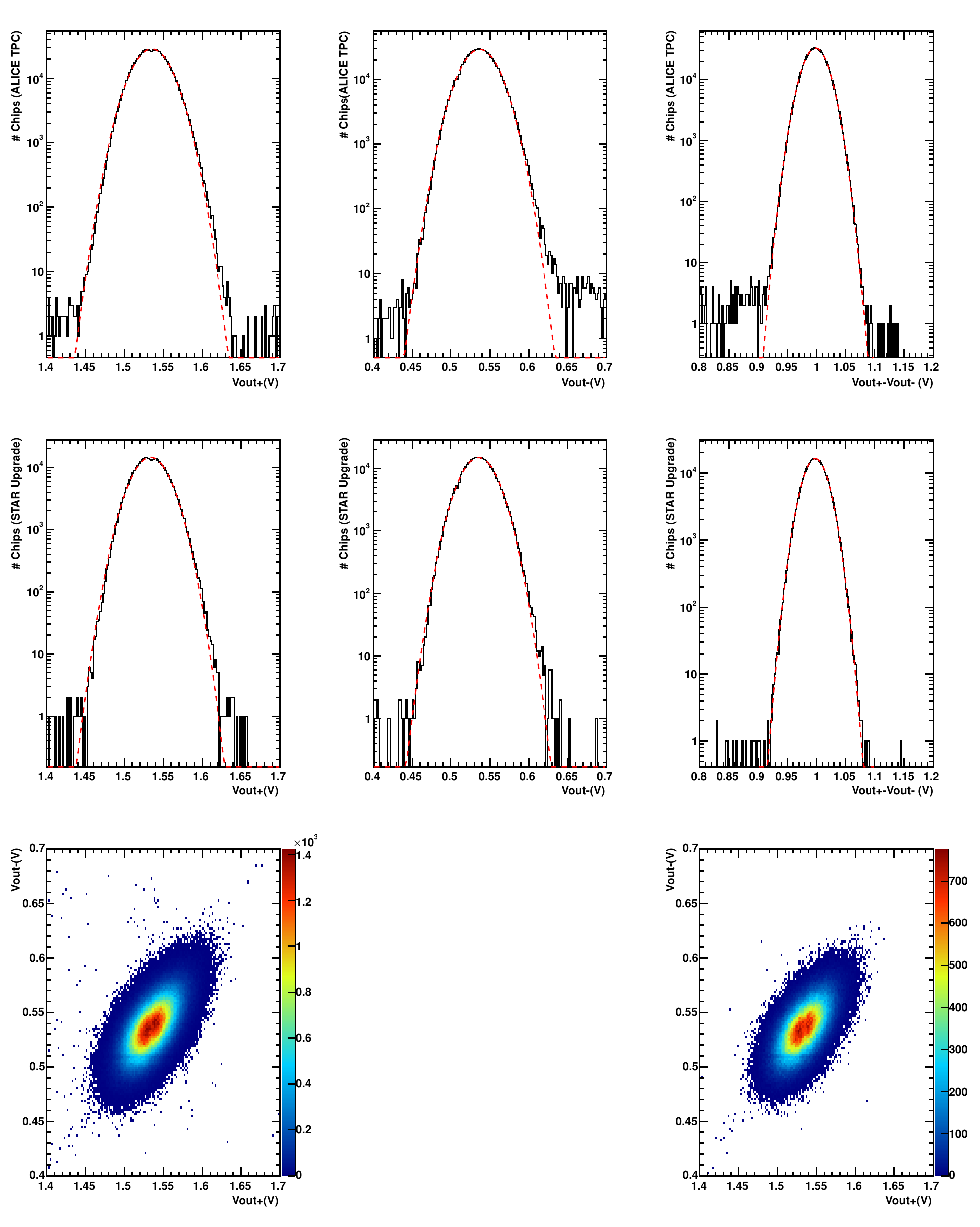}
\caption{The top and the middle rows show the distribution of VOUT+
  (left), VOUT- (middle) and VOUT+ - VOUT- (right) for the ALICE TPC
  PASA and for the STAR upgrade production, respectively. The dashed line shows a Gaussian fit to the distributions. In the bottom
  row the correlation of VOUT- versus VOUT+ is shown for ALICE TPC PASA
  (left) and STAR upgrade production (right).}
\label{fig:levelsall}
\end{center}
\end{figure}

The DC level of the circuit's differential output (baseline), which can be controlled by means of two external bias voltages VREFN and VREP, should be kept at a value sufficiently close to the bottom of the circuit's dynamic range, in order to preserve the maximum dynamic range. Since VREFP and VREFN will be common to all channels in one front-end card (or possible for all front-end cards in the whole detector), it is important to verify the dispersion of the circuit output baseline for given values  of the external bias voltages VREFP and VREFN. This is shown in Fig.~\ref{fig:levelsall}.
For the ALICE TPC PASA, these distributions are Gaussian like up to $\pm$4.2~$\sigma$, while for STAR upgrade the distribution of VOUT+, VOUT-, and VOUT+-VOUT- are Gaussian up to

 $\pm$5~$\sigma$. 
The number of channels
outside the same acceptance region as for the the ALICE TPC PASA,

In the bottom row left and right the correlation
between VOUT+ and VOUT- is shown for ALICE TPC PASA and STAR upgrade
PASA production respectively.
These plots show a very good correlation between the two output voltages.

Table~\ref{elect:yieldsPASA} and Table~\ref{elect:yieldsSTARPASA} show a summary of the most relevant results of the mass production test of the PASA for the ALICE TPC and the STAR upgrade, respectively. 
 \begin{table}
\begin{center}

\begin{tabular}{|l|cc|cc|}
\hline
Non accepted       & \multicolumn{4}{|c|}{ALICE frequency}    \\
 & \multicolumn{2}{|c|}{Channels} & \multicolumn{2}{|c|}{Chips} \\
                        & \# & \% &\# & \% \\
\hline
None               &     &   & 46\,585  &  98  \\
\hline 
Power (20\%)   &  & &   307 (180) &   0.65 \\
\hline
Conv.\ gain (8\%)  &  4922 (2880) &   0.65 &   914   &  1.9  \\
Peaking time (10\%) &  4492 (2880) &   0.59 &   673  & 1.4  \\

 VOUT+ (80 mV) & 2498 (1630) & 0.33& 654  & 1.4  \\
 VOUT -  (80 mV) & 2290 (1553) & 0.3&  593&  1.2  \\
 VOUT+  - VOUT - (80 mV) &2297 (1466) & 0.3&  633 & 1.3  \\  \hline
 Total amount of Chips & & & 1052 & 2.2\\
\hline
Total              & &&47\,637  & 100.0  \\
\hline
\end{tabular}
\caption{Acceptance levels for the PASAs as chosen for the ALICE TPC.}
\label{elect:yieldsPASA}

\end{center}
\end{table}

\begin{table}
\begin{center}

\begin{tabular}{|l|cc|cc|}
\hline
Non accepted       & \multicolumn{4}{|c|}{STAR frequency}    \\
 & \multicolumn{2}{|c|}{Channels} & \multicolumn{2}{|c|}{Chips} \\
                        & \# & \% &\# & \% \\
\hline
None               &     &   & 22\,609  &  99.2  \\
\hline
Power (20\%)   &  & &   63 (35) &   0.3 \\
\hline
Conv.\ gain (8\%)  &  2037 (560) &   0.55 &   160   & 0.7   \\
Peaking time (10\%) & 2014 (560) &   0.55 &  137   & 0.6  \\

 VOUT+ (80 mV) & 376  (231) & 0.10& 113 & 0.5  \\
 VOUT -  (80 mV) & 320  (205) & 0.09&101  &   0.4\\
 VOUT+  - VOUT - (80 mV) & 375  (228) & 0.10&111   & 0.5  \\  \hline
 Total amount of Chips & & & 181 & 0.80\\
\hline
Total              & &&22\,795  & 100.0  \\
\hline
\end{tabular}
\caption{Acceptance levels for the PASAs as chosen for the STAR TPC.}
\label{elect:yieldsSTARPASA}
\end{center}
\end{table}

Taken into account the acceptable region of operation, as discussed
above, a total of 1052 chips  were found to be outside this
region (mostly due to one channel). This gives a production yield for the ALICE TPC PASA of nearly 98\%. 
For the STAR upgrade production,  the number of chips outside the acceptance region, the same as for the ALICE TPC PASA, was 186. This corresponds to a yield of  99.2\%.

\subsection{System measurements}
\label{subsec:sys}

In this section we illustrate the performance of the PASA, with special focus on the noise, as measured  on the ALICE TPC fully instrumented with the electronics.

\begin{figure}
\begin{tabular}{cc}
\includegraphics[width=0.52\textwidth]{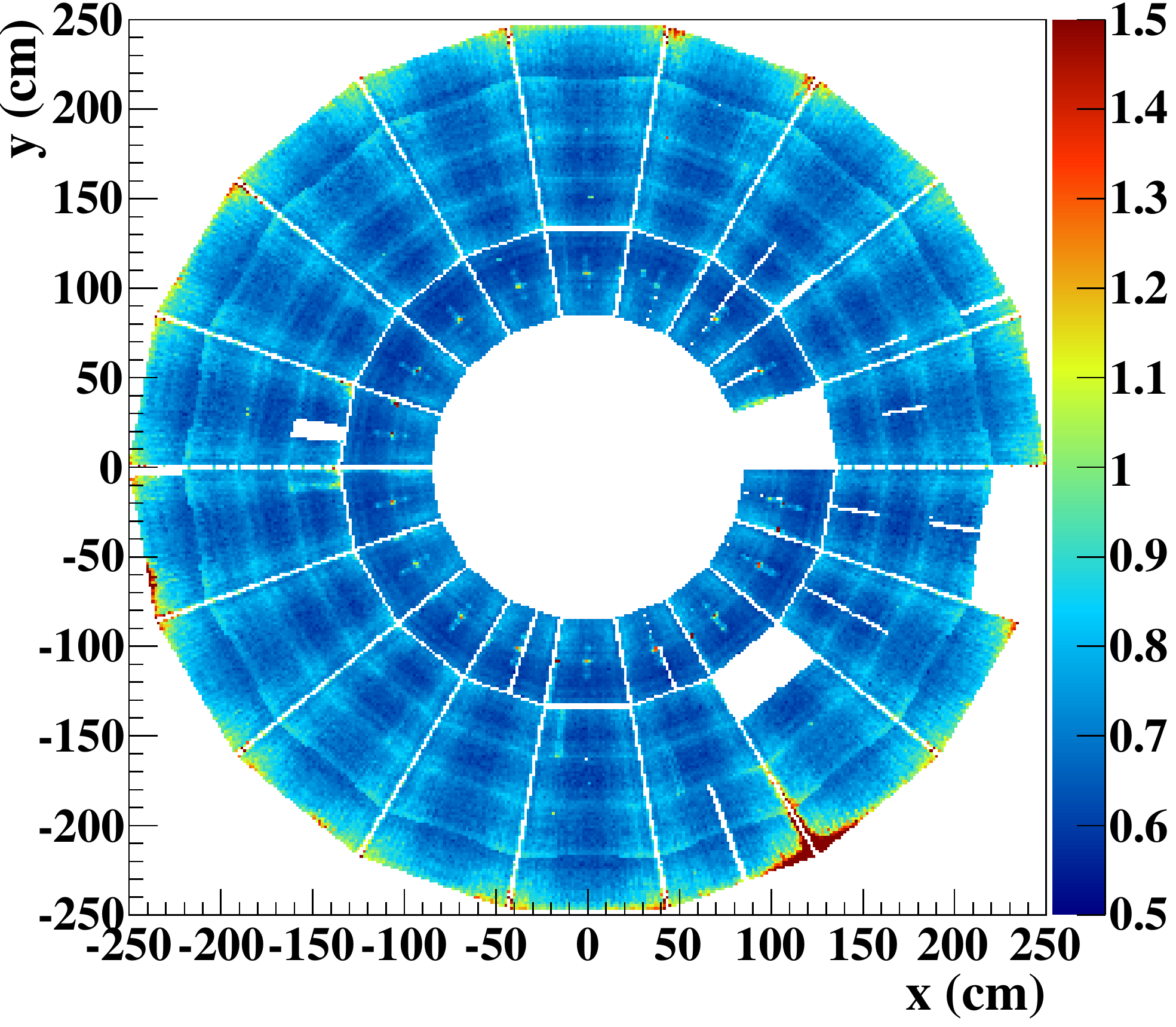}&
\includegraphics[width=0.52\textwidth]{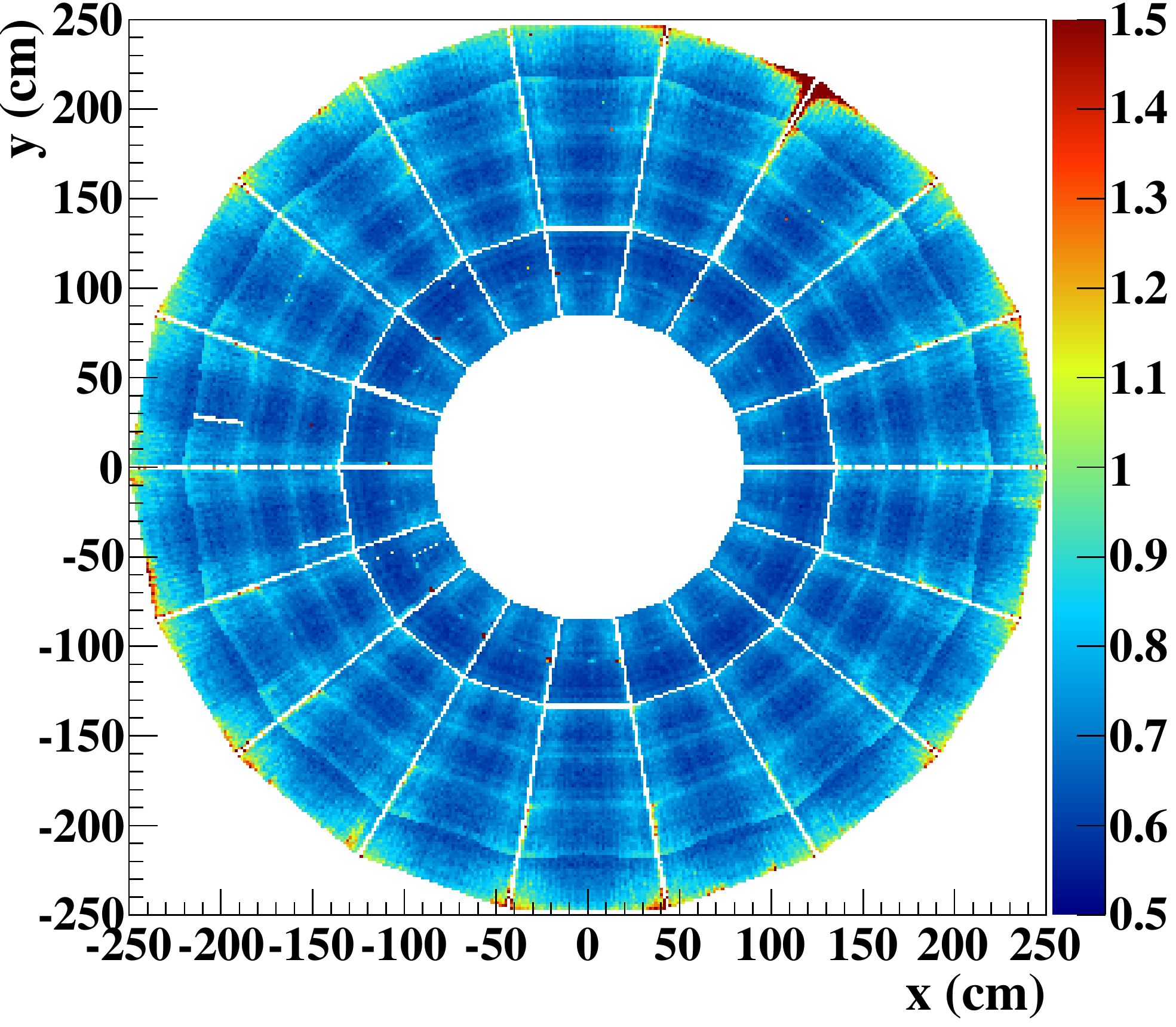}\\
\end{tabular}
\caption{Measured system noise for the A-side (left) and C-side
   (right) of the ALICE TPC detector \cite{Jens}.}
\label{fig:snoisexy}
\end{figure}

\begin{figure}
\begin{center}
\includegraphics[width=0.7\textwidth]{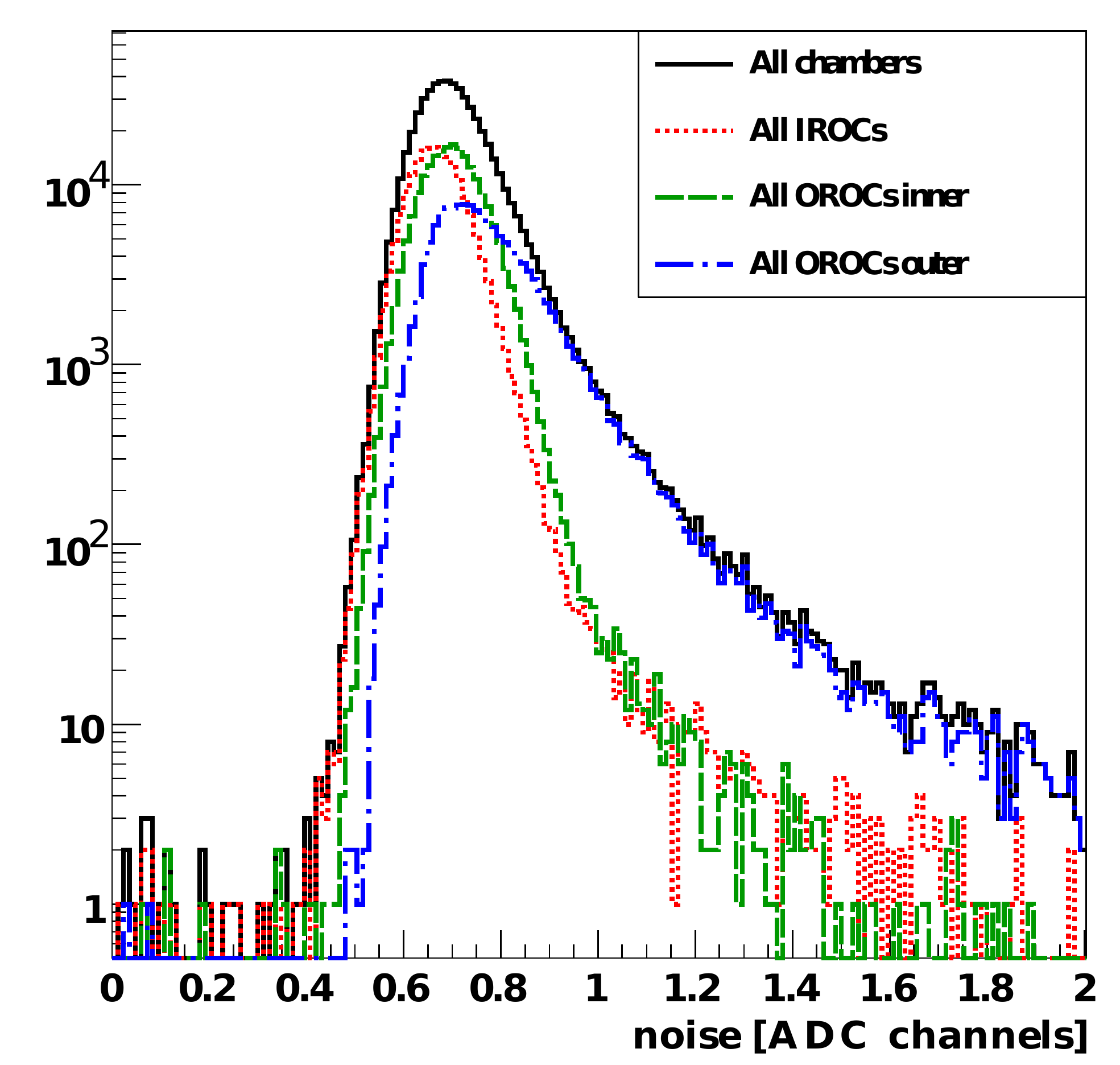}
\caption{The distribution of the overall system noise in the final
  setup in the ALICE experiment. The histogram for the three pad sizes
and the overall noise histogram are shown \cite{Jens}.}
\label{fig:snoisepad}
\end{center}
\end{figure}

Fig.~\ref{fig:snoisexy} shows the
measured system noise performance in ENC as a function of the x versus
y position in the A-side and in the C-side of the ALICE TPC
detector. As mentioned, each sector in the A-side and
C-side consist of three different pad sizes. The ENC distribution for each of the
three pad sizes separately and the overall noise distribution is shown in Fig.~\ref{fig:snoisepad}. 
The noise spread between the channels, which shows a very
systematic pattern is mainly due to the difference in trace length on the
chamber pad plane and to the different trace length on the front-end board.
Measured and extracted values given by the system measurement are summarized Tab.~\ref{tab:Measured}.

  \begin{table}[htb]
\begin{center}
\begin{tabular}{|l|c|c|c|}\hline
    Parameter                                                            & IROC       & OROC1  &  OROC2 \\ \hline \hline
    Trace length  (cm)                                               & 120          & 120         & 150\\ \hline
    ADC- counts/cm                                                   & 0.21         & 0.22         & 0.23\\ \hline
    Pad capacitance (pF)                                          & 6-10         & 6-10         & 6-10\\ \hline
    FEE capacitance (pF)                                          & 3-5           & 3-5           &  3-5\\ \hline
    Noise ADC-counts (without traces)                   & 0.589       & 0.603      & 0.624\\ \hline
    Noise ADC-counts (with traces)                        & 0.67          & 0.7          & 0.78\\ \hline
    Noise in electrons  (without traces)                   & 570          &  584         &    604 \\ \hline
    \hline
    Extracted capacitance (pF)                                  & 11-13      &  11-14     &  12-14 \\ \hline
    Extracted Average capacitance (pF)                  & 18            &  21           &  26 \\ \hline
    Extracted noise slope (1.3 pF/cm assumed)     & 15.6         & 15.2        & 15.9\\ \hline

  \end{tabular}
\end{center}
\caption{The seven first rows are values given by the system measurement  \cite{Jens}. The three last rows are extracted values.}. 
\label{tab:Measured}
\end{table}

An overall typical mean noise of 0.71~ADC counts was here achieved. 
This converts into an ENC equal to 710 electrons and is very close to optimum of what we can expect
from a system of this complexity. This value corresponds to an average system capacitance of about 21 pF.

\begin{figure}
\begin{center}
\includegraphics[width=0.8\textwidth]{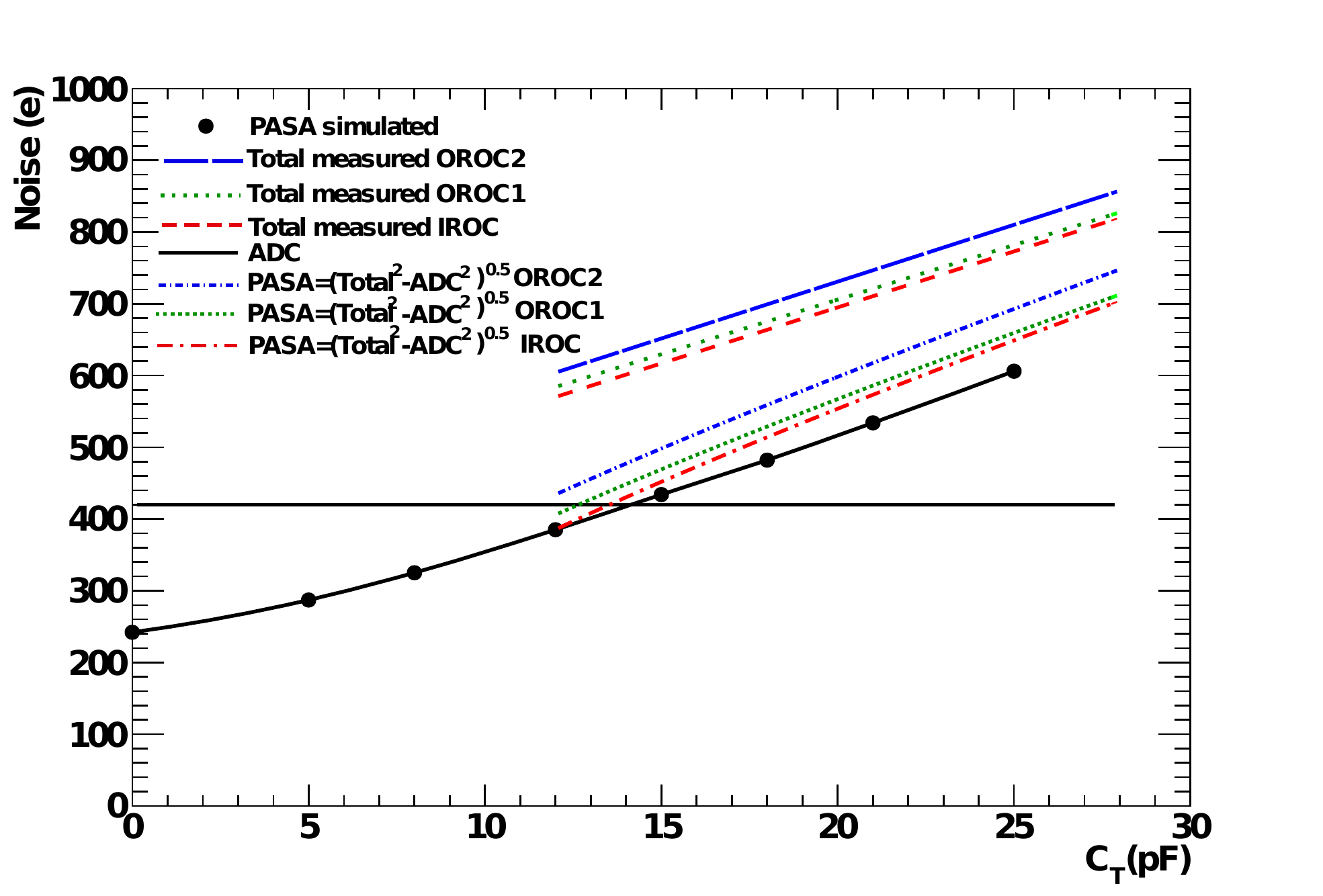}
\end{center}
\caption{Noise versus the total capacitance C$_{T}$. C$_{T}$ varies from 12 to 28 pF with a mean value around 21 pF.}
\label{fig:noiselines}
\end{figure}

The total measured system noise for the IROC, for the OROC1
and for the OROC2 as a function of the total detector capacitance is
shown in Fig.~\ref{fig:noiselines} (full-lines). After subtracting the
ADC noise contribution of 420e ( $PASA=\sqrt{total^2-ADC^2}$), the noise contribution from the ALICE TPC
PASA can be compared with the simulated one (dashed lines in Fig.~\ref{fig:noiselines}).
The measured noise is found to be, as expected, a little larger than
the simulated value.

\section{Conclusions}
\label{sec:conclu}

This paper describes the full custom design of the ALICE TPC
PASA. Simulation results of the layout with extracted parasitics done
prior to the submission of the chip have been presented. About 48000
chips, with a total of about 762 176 channels have been
fabricated and fully tested. Theoretical calculation, simulation and system
measurements results show very good agreement with each other.
The ALICE PASA production has successfully satisfied all specifications with good margin. 
In particular, the evaluation of
the noise in the fully commissioned  system shows an ENC noise behavior of
typically 0.71 ADC-counts  (710 electrons). This is close to optimum of what we can expect for
such a complex system including both PASA, ADC (ALTRO) and other
electronics placed in the vicinity of the ALICE TPC PASA. 

A baseline shift and a small offset in power consumption is
seen, due to manufacturing variations that will result in process and device parameter
variations from lot-to-lot, wafer-to-wafer, die-to-die, and device-to-device.

The extreme chip and system tests show a yield of about 98\%. Losses are mainly due
to one channel being outside the acceptance region. This very high yield is
the result of a robust design optimized using
extensive simulations, including extracted parasitics from
layout, for the recommended AMS corners,  
and of the techniques used in the
layout to reduce the effect of the process variations. 
The circuit was realized in AMS's C35B3C1 0.35 um CMOS process,
has a area of 18 mm$^2$ and the die is packed in a TQFP 144 package.

The ALICE TPC PASA has been selected for the upgrade at
STAR TPC (11500 chips), it also being used at
MIPP/Fermilab (1100 chips) and at BONUS/JLAB (a few hundred chips).

\section{Anknowledgement}

We would like to thank BMBF for the financial support of this project.
We also thank Marcus Dorn for his IT support and maintainance of the
Cadence system.
We also thank Kjetil Ullaland   for valuable comments to this manuscript.

\end{document}